\documentclass[prx,twocolumn,superscriptaddress,showpacs,notitlepage,floatfix,bibnotes]{revtex4-2}
\usepackage{amsmath,amsfonts, amssymb, amsthm, dsfont}
\usepackage{yfonts}
\usepackage{bm}
\usepackage{mathrsfs}
\usepackage{graphicx}
\usepackage{verbatim}
\usepackage{wasysym}
\usepackage{xcolor}
\usepackage{tikz}
\usepackage{physics}
\usepackage{floatrow}
\usepackage{diagbox}
\usepackage{footmisc}
\usepackage{ulem}
\usepackage[caption=false,label font={bf,normalsize}]{subfig}
\AtBeginDocument{\tabcolsep=7pt} 
\usepackage{booktabs}

\newcommand{\comments}[1]{}

\renewcommand{\cal}[1]{\mathcal{#1}}

\newcommand{\xzzx}{XZZX\ }
\newcommand{\xy}{XY\ }
\newcommand{\pixz}{\Pi_{XZ}}
\newcommand{\piyz}{\Pi_{YZ}}
\newcommand{\hyz}{H_{YZ}}
\newcommand{\Hadamard}{H}

\newlength{\fighskip} \fighskip=2pt
\newlength{\figvskip} \figvskip=3pt
\newcommand*{\figbox}[2]{{\def\figscale{#1}\def\arraystretch{0.8}\arraycolsep=0pt\begin{array}{c}\vbox{\vskip\figscale\figvskip\hbox{\hskip\figscale\fighskip\includegraphics[scale=\figscale]{#2}}}\end{array}}}

\makeatletter
\def\l@subsubsection#1#2{}
\makeatother

\definecolor{new}{rgb}{.08,.05,.8}

\usepackage[pdfpagelabels,pdftex,bookmarks,breaklinks]{hyperref}
\definecolor{darkblue}{RGB}{0,0,127} 
\definecolor{darkgreen}{RGB}{0,150,0}
\hypersetup{
	colorlinks, 
	linkcolor=darkblue, 
	citecolor=darkgreen, 
	filecolor=red, 
	urlcolor=blue,
  pdfauthor = {Arpit Dua, Aleksander Kubica, Grace M. Sommers, David Huse, Liang Jiang, Steve Flammia, Michael Gullans},
  pdftitle = {Clifford-deformed Surface Codes}
}

\begin{document}
\title{Clifford-deformed Surface Codes}  
\author{Arpit Dua}
\thanks{These authors contributed equally.}

\affiliation{Department of Physics and Institute for Quantum Information and Matter, California Institute of Technology, Pasadena, California 91125, USA}
 
\author{Aleksander Kubica}
\thanks{These authors contributed equally.}
 \affiliation{AWS Center for Quantum Computing, Pasadena, California 91125, USA}
 \affiliation{California Institute of Technology, Pasadena, California, 91125, USA}
\author{Liang Jiang}
    \affiliation{AWS Center for Quantum Computing, Pasadena, California 91125, USA}
    \affiliation{Pritzker School of Molecular Engineering, The University of Chicago, Illinois 60637, USA}
\author{Steven T. Flammia}
\affiliation{AWS Center for Quantum Computing, Pasadena, California 91125, USA}
\affiliation{California Institute of Technology, Pasadena, California, 91125, USA}

\author{Michael J. Gullans}
 \thanks{Correspondence: adua@caltech.edu, mgullans@umd.edu}
 \affiliation{Joint Center for Quantum Information and Computer Science, NIST/University of Maryland, College Park, Maryland 20742 USA}

\begin{abstract}
Various realizations of Kitaev's surface code perform surprisingly well for biased Pauli noise. Attracted by these potential gains, we study the performance of Clifford-deformed surface codes ({CDSCs}) obtained from the surface code by applying single-qubit Clifford operators. We first analyze {CDSCs} on the $3\times 3$ square lattice and find that, depending on the noise bias, their logical error rates can differ by orders of magnitude. To explain the observed behavior, we introduce the effective distance $d'$, which reduces to the standard distance for unbiased noise. To study CDSC performance in the thermodynamic limit, we focus on random \mbox{CDSCs}. Using the statistical mechanical mapping for quantum codes, we uncover a phase diagram that describes random CDSC families with $50\%$ threshold at infinite bias. In the high-threshold region, we further demonstrate that typical code realizations outperform the thresholds and subthreshold logical error rates, at finite bias, of the best-known translationally invariant codes. We demonstrate the practical relevance of these random CDSC families by constructing a translation-invariant CDSC belonging to a high-performance random CDSC family. We also show that our translation-invariant CDSC outperforms well-known translation-invariant CDSCs such as the XZZX and XY codes.\end{abstract}


\maketitle

\section{Introduction}
Optimization of quantum error-correcting (QEC) codes for realistic noise models is one of the essential steps towards reducing overhead in fault-tolerant quantum computing.
Among QEC codes, Kitaev's surface code~\cite{Kitaev2003,Bravyi1998,Dennis2002} and a multitude of its variants~\cite{Bombin2010,Yoder2017,Ultrahigh2018,Tailoring2019,XZZX2021} are perhaps the most experimentally feasible as they can be implemented in two-dimensional architectures with only nearest-neighbor interactions and offer reasonably high circuit noise thresholds.
Realistic noise, however, is likely to exhibit bias toward one type of error. For example, biased noise can be found or engineered in the architectures based on superconducting qubits~\cite{Pop2014}, trapped ions~\cite{Nigg_2014} and spin qubits~\cite{burkard2021,Watson_2018}.
Furthermore, noise bias can be preserved through careful engineering of fault-tolerant protocols or quantum gates \cite{Aliferis08,Aliferis_2009,Puri_2020}. 

Bias in realistic noise can be beneficial from the perspective of QEC.
Two prominent examples are the \xy and \xzzx surface codes~\cite{Ultrahigh2018,XZZX2021}.
While both are Clifford-equivalent to the CSS surface code, they both significantly outperform the latter in terms of the threshold and subthreshold scaling of the logical error rate for noise biased towards Pauli $Z$ errors~\cite{Ultrahigh2018,Tailoring2019,Tuckett2020,XZZX2021,Darmawan2021}.

\begin{figure}[t]
\centering
\includegraphics[width=0.7\columnwidth]{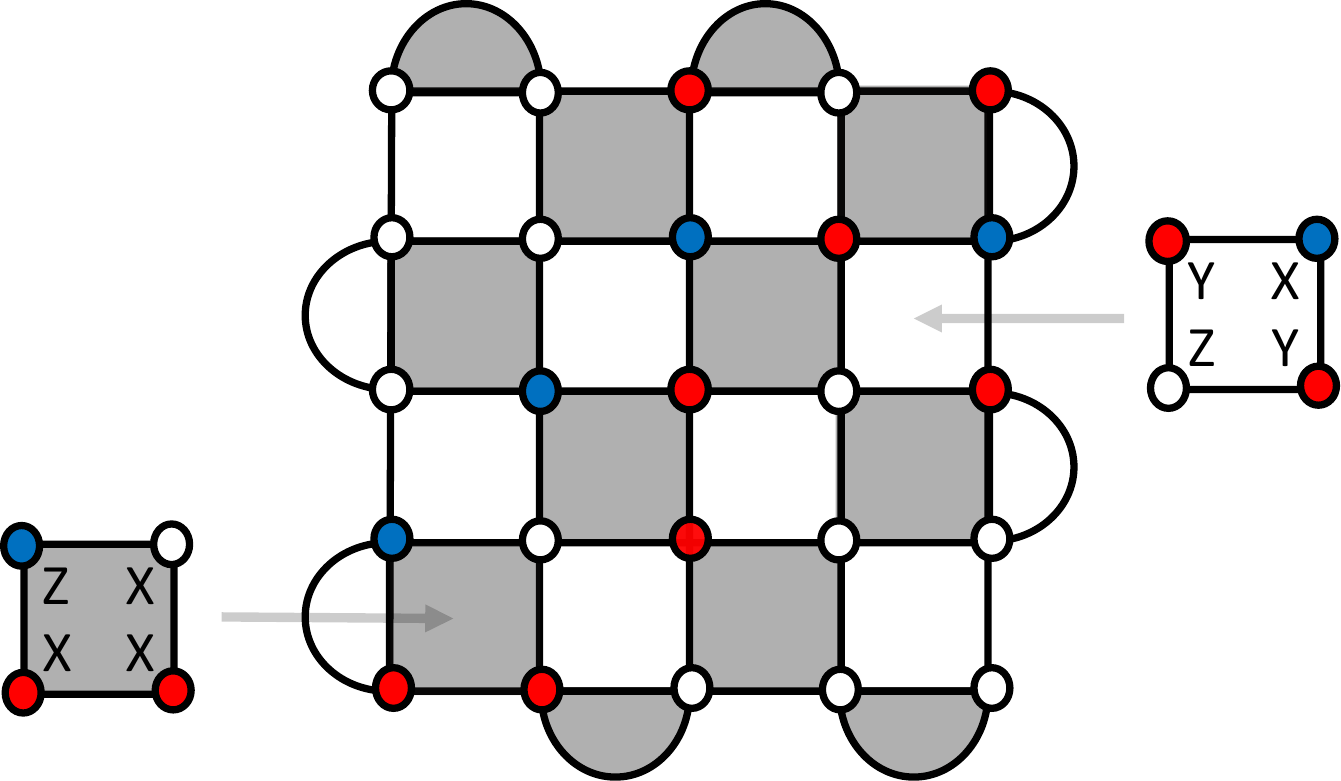}
\caption{A Clifford-deformed surface code on the $5\times 5$ square lattice with open boundary conditions.
Qubits are on vertices and, in the absence of Clifford deformations, $X$ and $Z$ stabilizer generators are associated with gray and white faces, respectively.
To modify the stabilizer group, we apply single-qubit Clifford operations $\Hadamard$ and $\Hadamard\sqrt{Z}\Hadamard$ to blue and red qubits, respectively.
We depict two stabilizer generators.
}
\label{fig:CDSCeg}
\end{figure} 

Inspired by these significant improvements, we provide a thorough study of the performance of surface codes against biased noise.
For concreteness, we focus on $L\times L$ square lattices, where $L$ is odd, with open boundary conditions; see Fig.~\ref{fig:CDSCeg}.
We consider stabilizer codes that are obtained from the surface code by the application of single-qubit Clifford operators. We collectively refer to such codes as Clifford-deformed surface codes (\mbox{CDSCs}).

We assume that every qubit is affected by independent and identically distributed (IID) biased Pauli noise with physical error rate  $p = p_X + p_Y + p_Z$ and bias $\eta\geq 0.5$, where $p_X = p_Y = p_Z /(2\eta)$ are the Pauli error rates.
Then, for a given error rate and bias, we seek \mbox{CDSCs} with optimal performance in terms of either the logical error rate or the code-capacity threshold.

For infinite bias, we explore the threshold phase diagram for random \mbox{CDSCs} and conjecture the existence of a connected region, where the corresponding codes have a threshold of $50\%$. 
For a moderate noise bias $\eta\sim 100$, we numerically find that typical random \mbox{CDSCs} in the high-threshold region can outperform the thresholds and subthreshold logical error rates of the best known translationally invariant codes, such as the \xy and \xzzx surface codes. 
We use the statistical-mechanical description of error correction thresholds \cite{Dennis2002,SMQECC2018} to optimize the QEC performance while also uncovering novel types of correlated percolation problems \cite{PCTbookStauffer1992,Coutinho20,Damavandi_2015} and phase transitions in disordered realizations of the 8-vertex model \cite{8VM_Fan_Wu_1970,8VM_Sutherland_1970}.
More generally, one can consider time-dependent constant-depth Clifford circuits to improve the QEC performance which shares similarities with
other dynamically-generated QEC codes~\cite{Brown13,Gullans21,Potter21,Hastings21} arising in the context of random circuits, measurement-induced phase transitions, and Floquet dynamics.


\section{Clifford deformations}
\label{sec:CDSCs}
Given any stabilizer code~\cite{gottesman1997stabilizer}, we can modify its stabilizer group by applying arbitrary single-qubit Clifford operators or other low-depth circuits \cite{vasmer2021morphing}.
Although the resulting code is topologically equivalent 
to the original code, the QEC performance can be improved, especially for biased Pauli noise.
This is exemplified by the \xzzx surface code~\cite{XZZX2021} that is obtained from the CSS surface code by applying the Hadamard gate $\Hadamard$ to every other qubit.

We are interested in the QEC performance of the \mbox{CDSCs} against Pauli noise biased towards Pauli $Z$ errors.
Since we assume a symmetry between Pauli $X$ and $Y$ errors, \textit{i.e.}, $p_X=p_Y$, it is sufficient to consider Clifford deformations whose action (by conjugation) on the single-qubit Pauli operators is either trivial, or $X\leftrightarrow Z$, or $Y\leftrightarrow Z$.
Such Clifford deformations are tensor products of the identity operator $I$, $\Hadamard$, and $\hyz = \Hadamard\sqrt{Z}\Hadamard$.
\vspace{-6mm}

\subsection*{Small CDSCs and effective distances}
\begin{figure}[t]
\centering
\sidesubfloat[]{\includegraphics[height=.15\textheight]{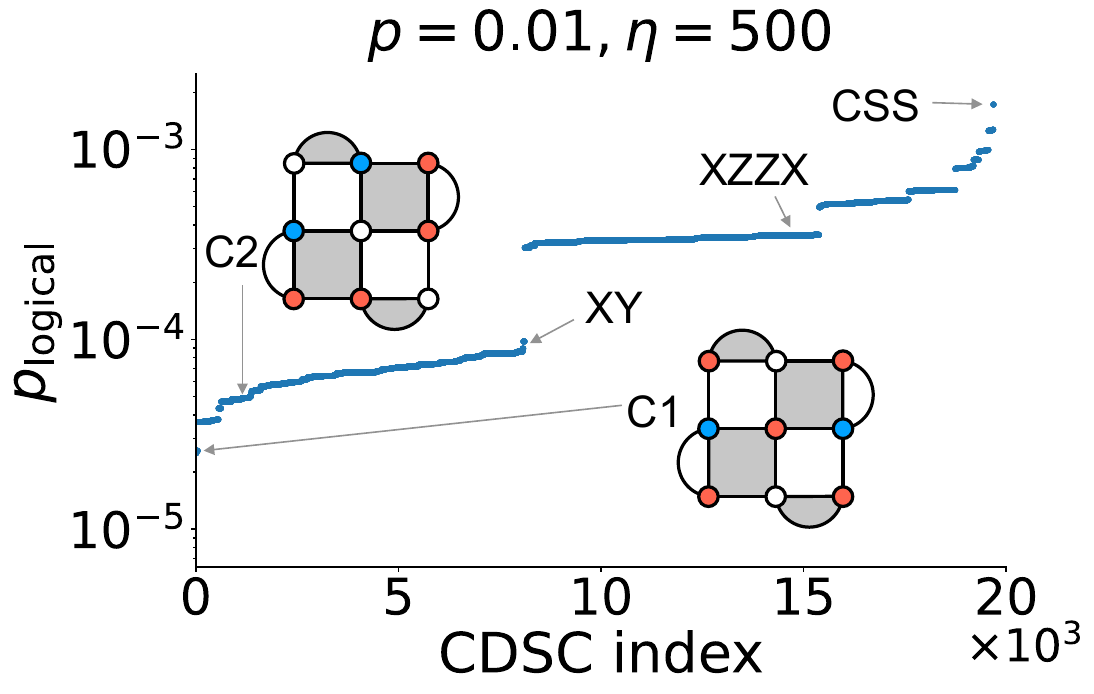}\hspace{9.2mm}}
\vspace{5mm}
\sidesubfloat[]{\includegraphics[height=.165\textheight, width=0.92\columnwidth]{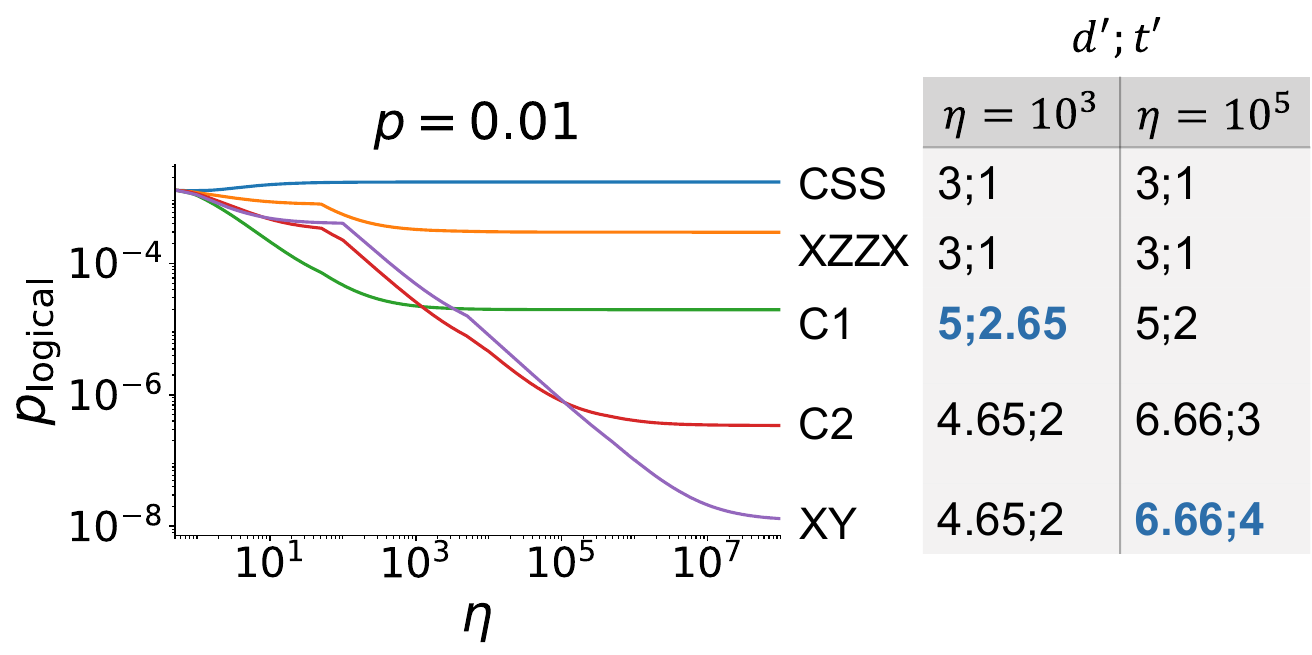}}
\caption{(a) Comparison of the logical error rate $p_\text{logical}$ for all possible \mbox{CDSCs} on the $3\times 3$ square lattice against biased Pauli noise with $p=0.01$ and $\eta = 500$.
(b) Logical error rate $p_\text{logical}$ as a function of the noise bias $\eta$ for selected \mbox{CDSCs} and fixed $p=0.01$. Effective distance $d'$ and half-distance $t'$ (up to three leading digits) are shown in the table.
For a given $\eta$, codes with the highest values $d';t'$ (colored in blue) perform best.
}
\label{3by3}
\end{figure}

We now study a small surface code and demonstrate that there is a wide range of possible Clifford deformations and associated code performances. For the representative values of the physical error rate
$p = 10^{-2}$ and bias $\eta = 500$ we compare the performance (measured in terms of the logical error rate) of \mbox{CDSCs} on the $3\times 3$ square lattice; Fig.~\ref{3by3}(a). 
There are $3^9 = 19683$ CDSCs to consider~\footnote{We could further reduce this number by incorporating the symmetries of the layout}.
Also, for a selection of codes we study their performance as a function of bias $\eta$ for fixed 
error rate $p=10^{-2}$; Fig.~\ref{3by3}(b).
For the depolarizing noise, that is a special case of Pauli biased noise with $\eta = 0.5$, all the \mbox{CDSCs} have the same performance, whereas in the regime of the large noise bias, \textit{i.e.}, $\eta \gtrsim 10^6$, the \xy surface code provides a clear advantage over other \mbox{CDSCs}, including the CSS and \xzzx surface codes.
To find the logical error rate we use the maximum-likelihood decoder.

Although all \mbox{CDSCs} on the $3\times 3$ square lattice have distance three, their logical error rates can be vastly different.
Moreover, the relative performance of codes changes with the noise bias; see Fig.~\ref{3by3}(b).
To explain this, we introduce notions of the \textit{effective distance} $d'$ and \textit{half-distance} $t'$ of a QEC code for the IID Pauli noise with error rate $p$ and bias $\eta$ as follows
\begin{eqnarray}
d' &=& \mathcal N^{-1} \log(p_\text{log} (1-p)^{-n}),\\
t' &=& \mathcal N^{-1}\log(p_\text{cor} (1-p)^{-n}).
\end{eqnarray}
Here $\mathcal N = \log (p\eta(1+\eta)^{-1}(1-p)^{-1})$ is a normalization factor, $n$ is the number of qubits, $p_\text{log}$ and $p_\text{cor}$ are the probabilities of the most likely Pauli operator that, respectively, implements a non-trivial logical operator and is non-correctable.
By definition, $t' \geq d'/2$.
Also, for the depolarizing noise,
we have $p_\text{log} = (p/3)^d (1-p)^{n-d}$ and $p_\text{cor} = (p/3)^t (1-p)^{n-t}$, where $d$ is the code distance and $t = \lceil(d-1)/2\rceil$, resulting in $d' = d$ and $t' = t$.

The quantities $d'$ and $t'$, roughly speaking, capture log probabilities of the most likely non-trivial logical operator and non-correctable error, respectively.
Since $d'$ and $t'$ depend on the noise bias $\eta$, they constitute a better proxy to the logical error rate than standard code distance $d$.
Note that this proxy is valid whenever $p$ is low and one can ignore entropic factors.
Knowing $d'$ and $t'$ allows us to predict which \mbox{CDSCs} on the $3\times 3$ square lattice perform best for given $p$ and $\eta$; see Fig.~\ref{3by3}(b). Note that we use $d'$ and $t'$ as alternative measures of code performance for our analysis of small codes. However, they are not efficiently calculable for large system sizes that we consider in the next section.

\section{Random CDSCs}
\label{sec:random_CDSCs}
We now consider the Clifford-deformed surface codes in the thermodynamic limit. In the thermodynamic limit, it is inefficient to study all possible Clifford deformations, even when the deformations are limited to single-qubit operations $I$, $H$ and $H_{YZ}$. Instead, we construct random realizations of the \mbox{CDSC} by selecting, independently for every qubit, $I$, $\Hadamard$, or $\hyz$ with probabilities $1-\pixz-\piyz$, $\pixz$, and $\piyz$, respectively. As discussed in Sec. II, we consider symmetric $Z$ bias ($p_X=p_Y$) and in this case, it is sufficient to consider these single-qubit Clifford operations to find inequivalent code performances of CDSCs.
Given $\pixz$ and $\piyz$, we refer to the resulting family of \mbox{CDSCs} as the $(\pixz,\piyz)$ random \mbox{CDSC}.
Our goal is to understand the performance of a typical realization of the $(\pixz,\piyz)$ random \mbox{CDSC} \cite{random}.

\begin{figure}[t]
\centering
\includegraphics[width=0.99\columnwidth]{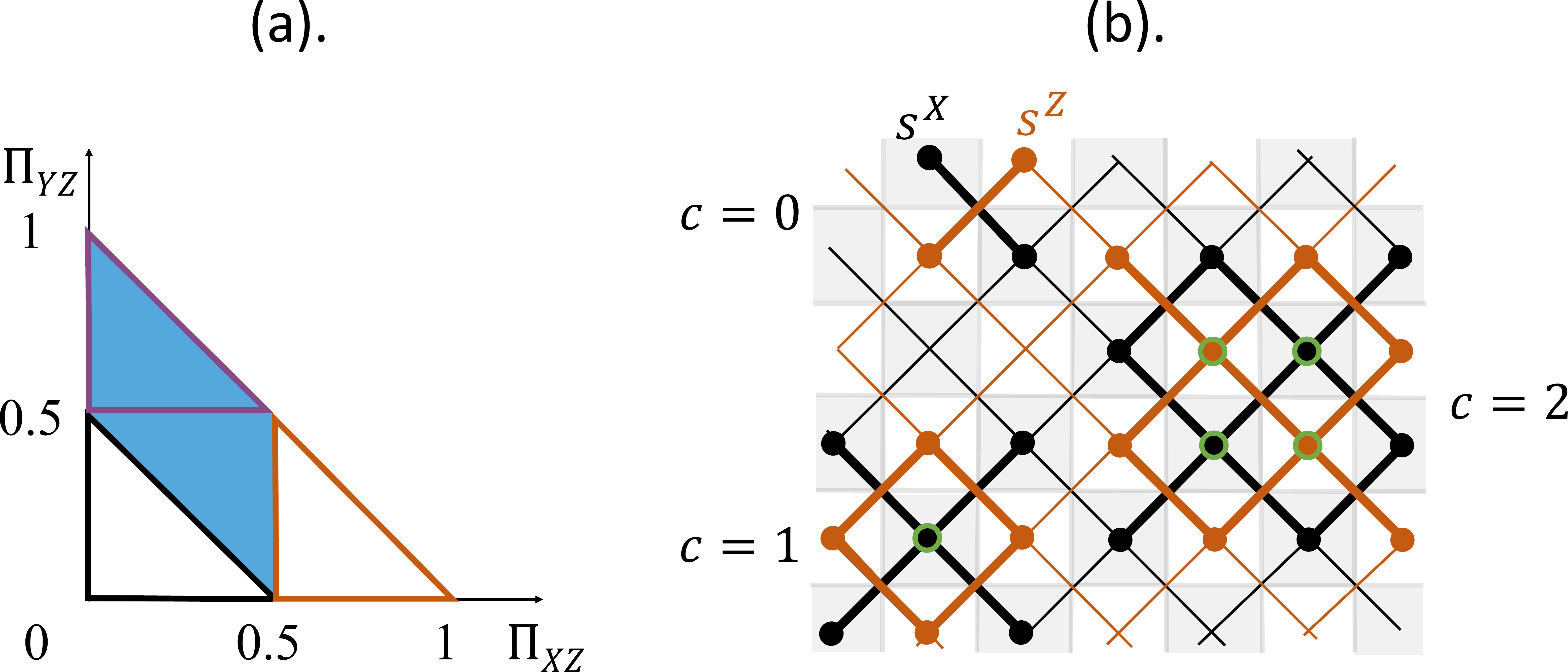}
\caption{(a) Phase diagram for random CDSCs at infinite bias.  Typical realizations of random CDSCs in the blue region of the $(\pixz,\piyz)$ phase space have a 50\% threshold.
The black and orange boundaries enclose regions where the $s^X$ and $s^Z$ sublattices order; in the region enclosed by purple boundaries we expect the system to order due to the $J_Y$ constraints. (b) Random-bond Ising model associated with a \mbox{CDSC}; the Hamiltonian is obtained by applying the deformations that define the \mbox{CDSC} to the Hamiltonian in Eq.~\eqref{eqn:SM_Hmain}. We highlight the clusters ($c=0,1,2$) used in the approximate analytical method to estimate thresholds for the XY code~\cite{supplemental}.
Spins circled in green are summed over while others are fixed. 
}
\label{fig:phase_random_stat_mech}
\end{figure}

To perform a systematic study of CDSC performance in the thermodynamic limit, we now explore the phase space $(\pixz,\piyz)$ of all random \mbox{CDSCs}, where, by definition, $\pixz,\piyz\geq 0$ and $\pixz + \piyz \leq 1$; see Fig.~\ref{fig:phase_random_stat_mech}(a). The $(0,0)$ and $(0,1)$ random \mbox{CDSCs} correspond to the CSS and XY surface codes, respectively.
Due to the code and noise symmetries, the $(\pixz,0)$ and $(1-\pixz,0)$ random \mbox{CDSCs} are equivalent.
We choose the $(0,0)$ random CDSC to be the reference code to which we apply Clifford deformations. Note that the XZZX code is a realization of the $(0.5,0)$ random \mbox{CDSC}. 

\subsection{Infinite bias phase diagram for random CDSCs}To evaluate the performance of random \mbox{CDSCs}, we first study their behavior at infinite bias. In particular, using the principles of statistical mechanics, 
we conjecture the existence of a region in the $(\pixz,\piyz)$ phase space where the corresponding random \mbox{CDSCs} have a $50\%$ threshold; see Fig.~3(a).
We support this conjecture via tensor network simulations for random \mbox{CDSCs} by applying a straightforward adaptation of the tensor network decoder by Bravyi et al.~\cite{BSV2014}, which efficiently approximates the maximum likelihood decoder~\cite{supplemental}.

To understand the $(\Pi_{XZ},\Pi_{YZ})$ phase space, we invoke the connection between thresholds of QEC codes and phase transitions in statistical-mechanical classical spin models~\cite{Dennis2002,Wang2003,Bombin2009,Depolarizing2012,kovalev2014spin,Kubica2018,SMQECC2018}.
The key idea behind this statistical-mechanical mapping is that a certain partition function of the disordered statistical-mechanical model gives the probability of a logical class of errors. Due to this, the mapping relates the optimal error-correction threshold of \mbox{CDSCs} to a critical point along the Nishimori line~\cite{Nishimori_1981}, which is a symmetric submanifold in the parameter space,
for a two-dimensional random-bond Ising model (RBIM)\footnote{Here, we use the nomenclature random-bond Ising model (RBIM) for a general disordered statistical-mechanical model of Ising spins.}.
For the CSS surface code, which corresponds to the point $(0,0)$, this RBIM is given by the 8-vertex model~\cite{8VM_Fan_Wu_1970,8VM_Sutherland_1970} whose disordered Hamiltonian (neglecting boundary terms), contains the following two- and four-body terms, \textit{i.e.},
\begin{equation}
\mathcal H_E = -\sum_{\figbox{.12}{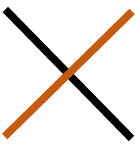}}
\tau_X J_X\figbox{.2}{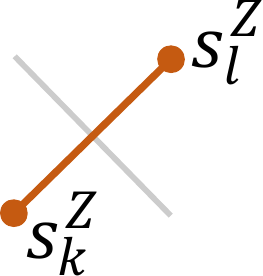}
+\tau_Y J_Y\figbox{.2}{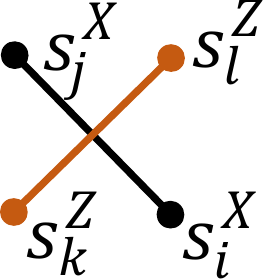}
+ \tau_Z J_Z\figbox{.2}{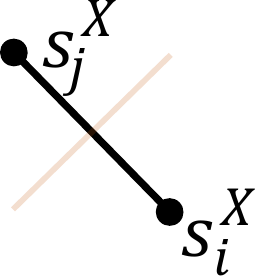},
\label{eqn:SM_Hmain}
\end{equation}
where $s_*^X,s_*^Z = \pm 1$ are Ising spins associated with every $X$ and $Z$ stabilizer generator of the surface code, and the summation is over locations of qubits, \textit{i.e.}, all $\figbox{.12}{eq_summand}$ and $\figbox{.12}{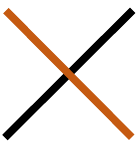}$ crossings; see Fig.~3(b).
$E$ is a Pauli error for the CSS code that determines the quenched disorder $\tau_P = \pm 1$ via $EP = \tau_P P E$ for every single-qubit Pauli $P=X,Y,Z$.
Moreover, $J_P= \frac{1}{4\beta}\log\frac{p_P^2(1-p)}{p_Xp_Yp_Z}$, where $\beta$ is the inverse temperature, is the coupling strength along the Nishimori line.
For the \mbox{CDSC}, depending on the deformation $\Hadamard$ or $\hyz$,
the relevant Hamiltonian is obtained from Eq.~\eqref{eqn:SM_Hmain} by permuting the corresponding coupling constants $\tau_{X}J_{X}\leftrightarrow \tau_{Z}J_{Z}$ or $\tau_{Y}J_{Y}\leftrightarrow \tau_{Z}J_{Z}$, respectively. 

Since for $\eta = \infty$ there are only Pauli $Z$ errors, the disorder for the coupling constant $J_Z$ in Eq.~\eqref{eqn:SM_Hmain} is fixed, \textit{i.e.}, $\tau_Z = 1$ and $J_Z =\infty$. Thus, the product of any spins that are coupled by $J_Z$ has to be $1$. Moreover, since $p_X=p_Y=0$, $J_X=J_Y$. The constraints are specified, according to the Clifford deformation, as follows,
\begin{equation}
I: \figbox{.195}{eq_sxx} = 1, \quad
\Hadamard: \figbox{.195}{eq_szz} = 1, \quad
\hyz: \figbox{.195}{eq_sxxzz} = 1,
\label{eqn:constraints}
\end{equation}
which we call $J_Z$, $J_X$ and $J_Y$ constraints, respectively. Note that, in the statistical-mechanical mapping, the stabilizers that are violated due to a $Z$ error on a qubit correspond to the spins involved in the constraint associated with that qubit. 

We first focus on the point $(0.5,0)$, which is an extreme point of the 50\% threshold region.
To understand it, we start from the $(0,0)$ point and move toward $(0.5,0)$ along the $\pixz$ axis by applying $\Hadamard$ deformations.
Since $\Hadamard$ permutes the coupling constants $\tau_X J_X \leftrightarrow \tau_Z J_Z$, we get a correlated bond percolation problem for the $J_X$ and $J_Z$ constraints on the $s^Z$ and $s^X$ sublattices, respectively. To explain this concretely, we first state the regular bond percolation problem on the square lattice. Each bond on the square lattice is assigned a probability of occupancy. If this probability of occupying a bond is less than the critical value of 0.5, the probability of finding a macroscopic connected cluster of occupied bonds is zero in the thermodynamic limit of infinite system size. In our case, we have two square sublattices that are dual to each other (see Fig. 3(b)) and we are interested in the probability of a macroscopic connected cluster of occupied bonds on both of these square sublattices. The Hadamard $H$ gate turns an occupied bond/constraint on one lattice to an occupied bond/constraint on the dual lattice associated with the same qubit. Hence, the bond percolation problem on the two dual square sublattices is correlated.


For $\pixz<0.5$, \textit{i.e.}, below the critical point for the bond percolation problem, the $J_X$ constraints do not percolate. However for $\pixz=0.5$, both the $J_Z$ and $J_X$ constraints exhibit critical fluctuations. This is indicative of a second-order phase transition. Generally speaking, as we approach a second-order phase transition, the correlations transition from exponential to power-law decays and right at the critical point, we get scale-invariant fluctuations extending over the full system.
For our case, these critical fluctuations manifest in the probability of finding a connected cluster of weight $A$ of $J_X$ and $J_Z$ constraints on the $s^Z$ and $s^X$ sublattices, respectively. This probability scales as $\mathcal{O}(1/A^{\tau})$, where $\tau=187/91$ is the Fisher exponent, while the weight of the largest cluster in the system scales as $L^{a}$ for $a=2/(\tau-1)$~\cite{PCTbookStauffer1992}.
There is also an important relation between the perimeter of the critical clusters $P$ and their area $A \approx P^{4 a/7}$ \cite{Saleur87}.

We present a heuristic argument that these critical clusters of constraints (that correspond to infinite strength couplings) are sufficient to order the entire system.
Assume we are in the ordered phase and take a droplet, \textit{i.e.}, a simply connected region of perimeter $\ell$ and area $A_\ell \ge \ell$.  Fixing $\epsilon\in(0,0.75)$, the probability $\Pi_\ell$ that this droplet has an overlap with at least one critical cluster of perimeter $P \ge \mathcal{O}(\ell^{1+\epsilon})$ scales as $\Pi_\ell \approx A_\ell/\ell^{(1+\epsilon) 4a (\tau-2)/7} \ge O(\ell^{1-(1+\epsilon)5/84})$.  Thus, $\Pi_\ell$ converges to one with increasing $\ell.$
Due to the spin constraints, reversing the spins in such a droplet also requires flipping \textit{all} the spins in the overlapping  critical clusters, thereby, leading to an  energetic cost, $\beta \Delta \mathcal {H}_E \ge \mathcal{O}(\ell^{1+\epsilon})$ for typical disorder realizations along the Nishimori line; since this is a 2D RBIM, the energy cost scales with the perimeter and for the overlapping clusters, it is dominated by their combined perimeter instead of the droplet perimeter $\ell$.
Applying Peierls' argument \cite{Griffiths64}, the energetic cost of flipping the spins in these droplets will typically dominate the entropic contributions to the free energy arising from the number of droplets  $N_\ell < 4^\ell$ with perimeter $\ell$.  
As a result, at any finite temperature, droplets with large perimeters will be prevented from fluctuating (flipping the spins), because of the critical clusters of spin constraints. 
Thus, the ordered phase along the Nishimori line is stable against the proliferation of macroscopic droplets of area $\mathcal{O}(L^2)$ for any finite temperature as $L \to \infty$.
Based on these arguments, we conjecture that typical $(0.5,0)$ code realizations remain ordered at any finite temperature along the Nishimori line, which implies that the $(0.5,0)$ random CDSC threshold is 50\%.
Numerical simulations confirm our predictions~\cite{supplemental}.  

We note that connected clusters of constraints that span the lattice imply the existence of a logical operator made of Pauli $Z$'s supported on the same qubits as the cluster constraints. In other words, the connected critical clusters identify non-local operators that commute with all local stabilizer terms. In particular, the critical percolation clusters on one of the two (dual) sublattices corresponds to the $Z$-logical operator. In two dimensions, numerical estimates of the typical length of the minimum spanning path indicate the scaling $\cal{O}(L^{1.1})$~\cite{PCTbookStauffer1992}.
This implies that the $Z$ distance (defined as the weight of the minimum-weight logical operator comprising Pauli $Z$'s) for the $(0.5,0)$ CDSC scales as $\cal{O}(L^{1.1})$.

To explain the region $\piyz\geq 0.5$, we recall that the $(0,1)$ CDSC is the XY code which has a 50\% threshold~\cite{Tailoring2019} and its associated RBIM has $J_Y$ constraints on all qubit locations.
Thus, for the XY code, the free energy cost of insertion of a disorder domain wall corresponding to a logical operator comprising Pauli $Z$’s in the RBIM diverges with $L$ for any finite temperature. The XY code is the limiting case of $\Pi_{YZ}=1$ in which the cluster of $J_Y$ constraints spans the entire lattice. More generally, we expect that a macroscopic cluster of $J_Y$ constraints is sufficient to order the entire system. Again, using the result of square lattice bond percolation having a critical point of 0.5, we expect macroscopic clusters of $J_Y$ constraints to occur for $\Pi_{YZ}\geq 0.5$. Furthermore, we argue that the macroscopic connected cluster of $J_Y$ constraints in the phase space region defined by $\piyz\geq 0.5$ is sufficient to order the system. 

In the region defined by boundaries $\pixz+\piyz > 0.5$ and $\pixz, \piyz< 0.5$, we cannot focus on the $J_Y$ constraints or $J_X$/$J_Z$ constraints alone. In fact, in this region, the $J_Y$ constraints combine with the $J_X$ and $J_Z$ constraints to form a macroscopic cluster of constraints on the $s^Z$ and $s^X$ sublattices, respectively. We conjecture that such a ``mixed" cluster can also stabilize the ordered phase i.e., ordering via these mixed macroscopic clusters could explain the interior of the phase diagram. 
An ordering arising from such a mixed cluster has not been studied in the field of percolation theory and arises especially for our problem of CDSCs under biased noise. We leave detailed investigations of this phenomenon to future work.

\subsection{Finite-bias performance of random CDSCs}
\begin{figure}[t!]
\centering
\includegraphics[width=\columnwidth]{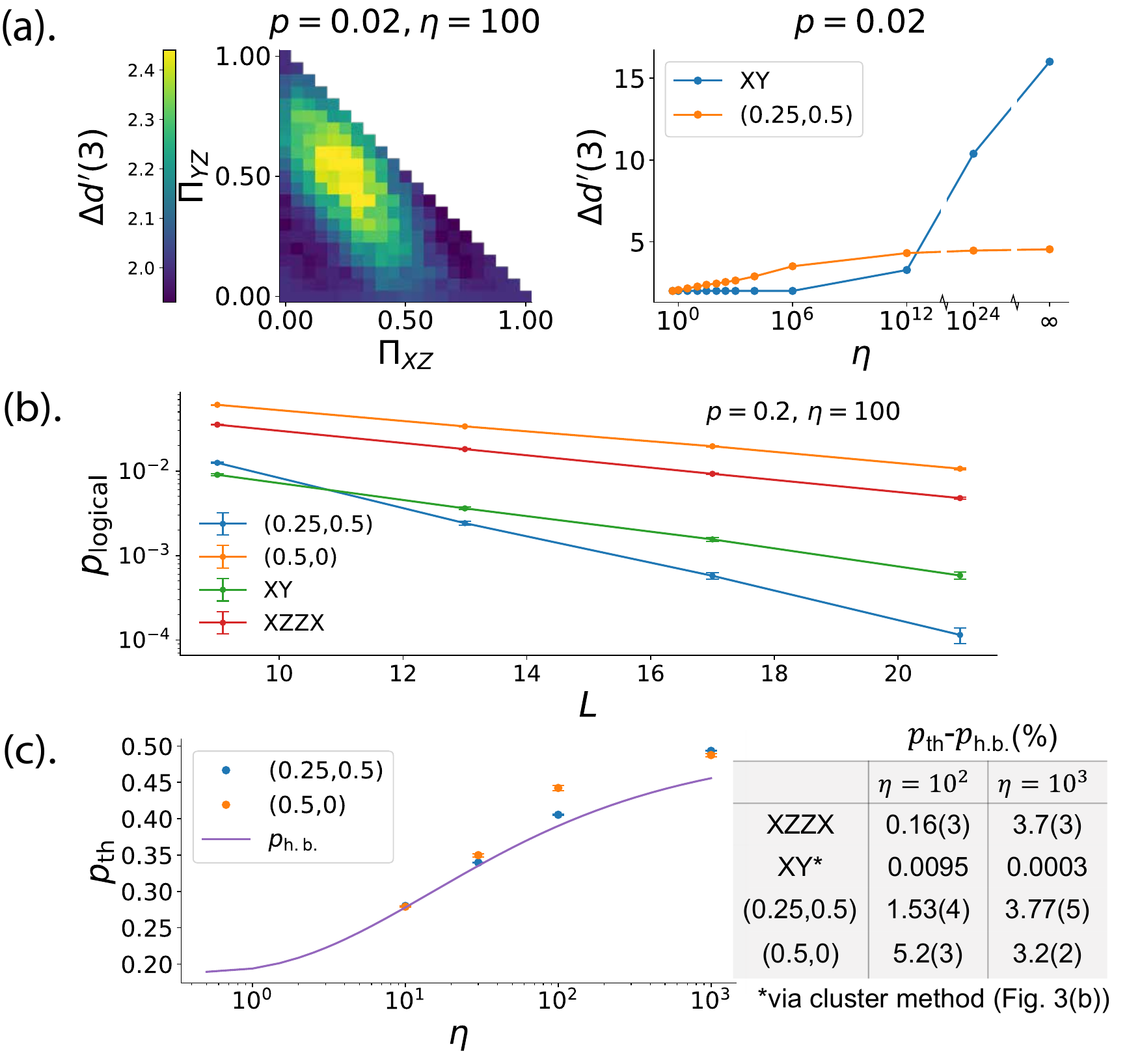}
\caption{(a) Effective distance increment $\Delta d^\prime(L=3)$ for random \mbox{CDSCs} and $p=0.02$, where we averaged over 2000 realizations for each point. (left) For $\eta=100$, the highest value is obtained for the $(0.25,0.5)$ random \mbox{CDSC}.
(right) $\Delta d^\prime(3)$ as a function of $\eta$.
For large $\eta$, $\Delta d^\prime (3)$ for the XY surface code starts to increase with $\eta$ (which is consistent with the $Z$ distance of the XY surface code being $L^2$ for infinite bias~\cite{Tailoring2019}) and exceeds $\Delta d^\prime (3)$ for the $(0.25,0.5)$ random \mbox{CDSC}.
(b) Subthreshold logical error rates $p_\text{logical}$ of random \mbox{CDSCs} on the $L\times L$ square lattice for  $p= 0.2$ and  $\eta = 100$.
(c) Thresholds $p_\text{th}$ for the $(0.25,0.5)$ and $(0.5,0)$ random \mbox{CDSCs} and hashing bound $p_{\text{h.b.}}$ as functions of $\eta$. The difference  $p_\text{th}-p_{\text{h.b.}}$ is tabulated in percentage for some codes and biases.}
\label{fig:dprime_random}
\end{figure}

Moving away from infinite bias, we want to find $(\pixz,\piyz)$ random \mbox{CDSCs} that perform best for a given error rate $p$ and bias $\eta$.
As a proxy to the logical error rate, we use the effective distance $d'$ and half-distance $t'$. 
For instance, for the XY surface code $d'(L) = L - \mathcal N^{-1}\log(2\eta)$ in the regime of low $p$ and moderate $\eta$~\footnote{The most likely logical operator is a string along the boundary, consisting of $L-1$ $Z$’s and one $X$ or $Y$}. 
Since we do not know efficient methods of calculating $d'$ and $t'$ for an arbitrary \mbox{CDSC}, we consider small linear system sizes $L$ and study the effective distance increment $\Delta d^\prime (L) = d'(L+2) - d'(L)$ to evaluate code performance. We expect $\Delta d^\prime(L)$ to be indicative of random \mbox{CDSCs} with good sub-threshold scaling; see Fig.~\ref{fig:dprime_random}(a).

Using tensor network simulations, we find that the $(0.25,0.5)$ random \mbox{CDSC}, which is one of the best performing code families according to $\Delta d'(3)$, outperforms the XY and XZZX surface codes in terms of subthreshold scaling of the logical error rate; see Fig.~\ref{fig:dprime_random}(b) and Supplemental Material~\cite{supplemental}.
Moreover, its threshold exceeds the hashing bound and the thresholds of the XZZX and XY codes at a moderate bias of $\eta = 100$; see Fig.~\ref{fig:dprime_random}(c).

We remark that the statistical-mechanical mapping for the XY surface code without disorder is self-dual whenever $p_X=p_Y$.
Hence, one can use the cluster methods~\cite{Nishimori_Ohzeki_Nihat_2008,Ohzeki_2009} with the clusters in Fig.~\ref{fig:phase_random_stat_mech}(b) to estimate thresholds at any noise bias $\eta$.
We describe the method in Supplemental Material~\cite{Note1} and show the results for two biases in Fig.~\ref{fig:dprime_random}(c). 
Extending these methods to the non-self-dual regimes at a finite bias (which would apply to other \mbox{CDSCs}) remains an outstanding challenge.

\begin{figure}
    \centering
    \includegraphics[scale=0.1]{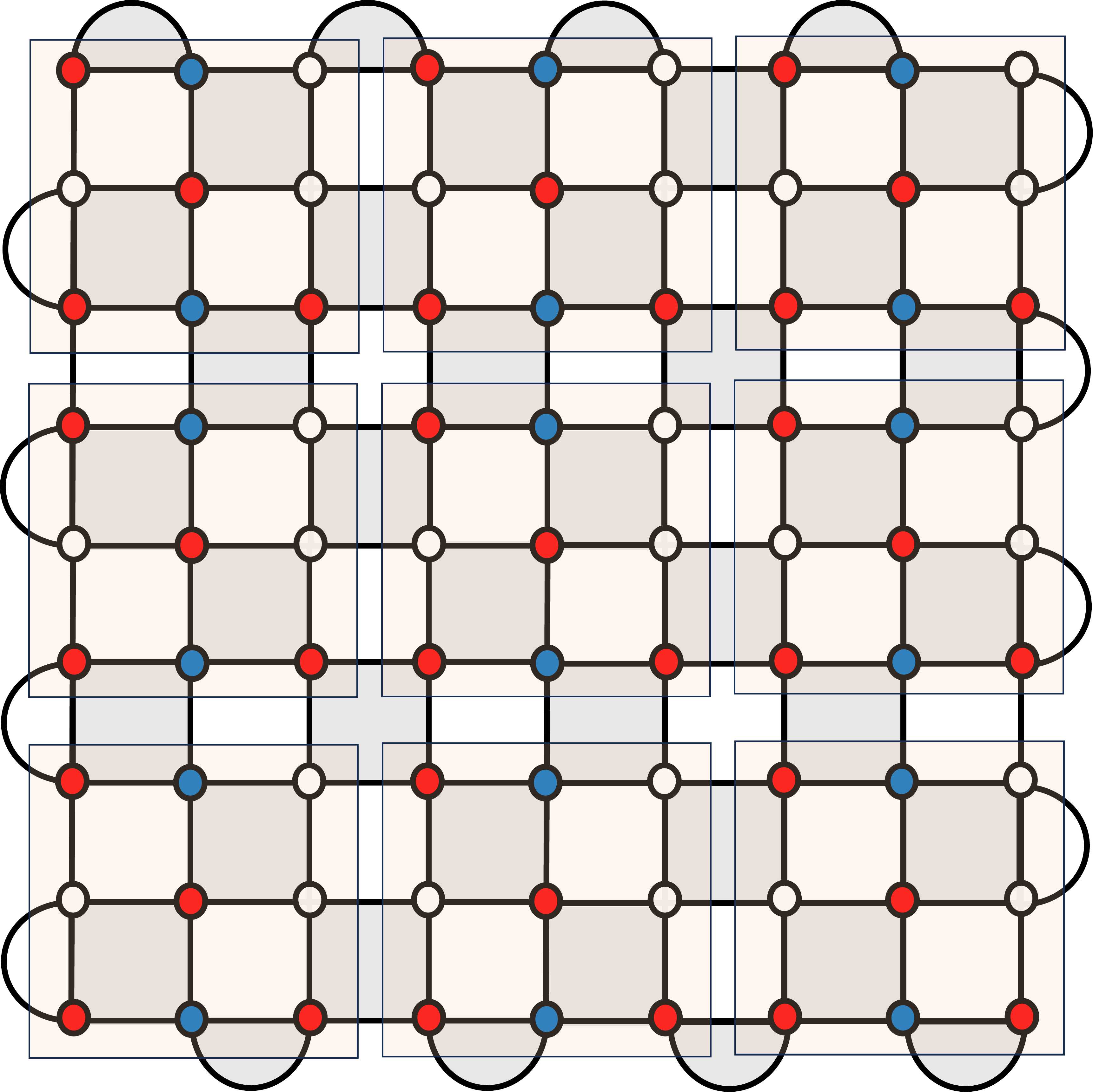}
    \caption{A translation-invariant CDSC belonging to the (2/9,4/9) CDSC family. The unit cell of the deformation is highlighted using a square patch. Within the unit cell, the number of $\Hadamard$s applied (qubits highlighted in blue) is half the number of $\Hadamard\sqrt{Z}\Hadamard$'s (qubits highlighted in red).}
    \label{fig:TI_code}
\end{figure}

\section{Translation-invariant codes from random CDSC families}
\label{sec:TI_CDSC}
Random CDSC families, investigated in the previous section provide guidance on designing high-performance translation-invariant CDSC. In particular, they suggest ratios of Hadamard $\Hadamard$ and $\Hadamard\sqrt{Z}\Hadamard$ that could be applied in the unit cell associated with a translation invariant Clifford deformation. In Sec.~\ref{sec:random_CDSCs}, we found that the $(0.25,0.5)$ random CDSC family performs the best among the random CDSC families in terms of subthreshold failure rates. Hence, as an example, we consider a translation-invariant realization of random CDSC family $(2/9,4/9)$, which is close in the $(\Pi_{XZ},\Pi_{YZ})$ phase space to the $(0.25,0.5)$ family. We choose this family $(2/9,4/9)$ instead of $(0.25,0.5)$ since we consider surface codes on $L\times L$ square lattices, where $L$ is odd. 

\begin{figure}
    \centering
    \includegraphics[scale=0.45]{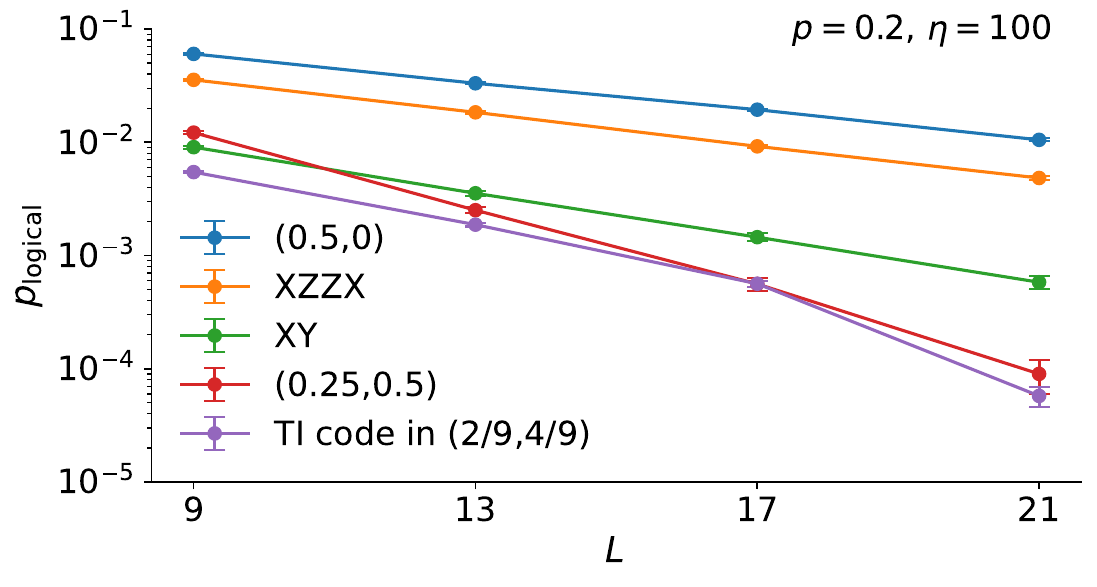}
      \includegraphics[scale=0.45]{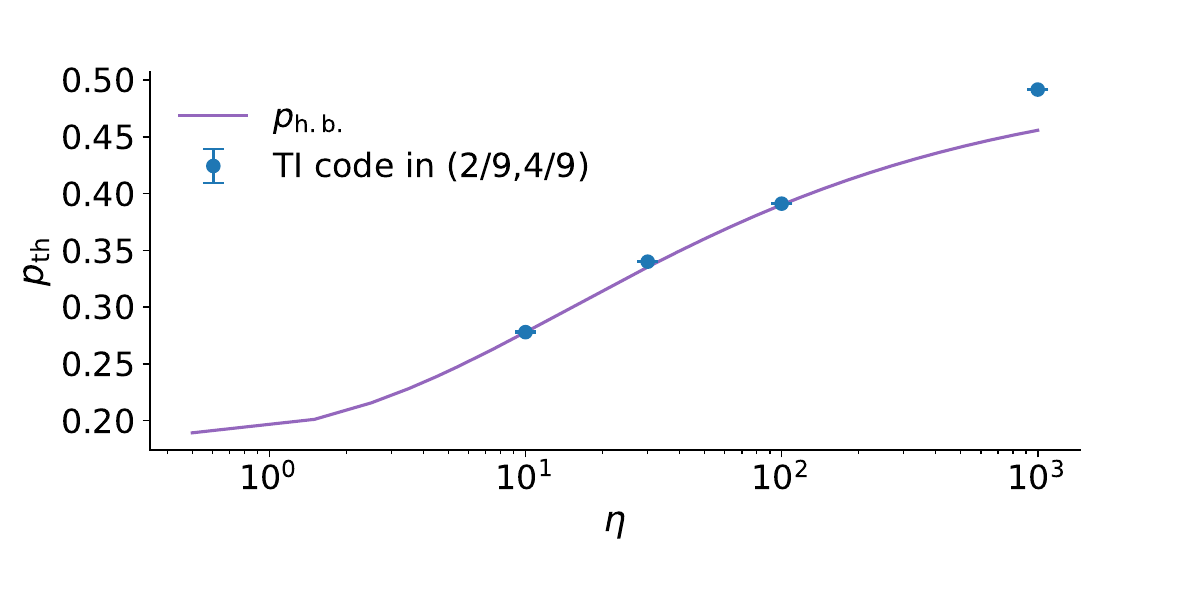}
    \caption{Top: Subthreshold logical error rates $p_\text{logical}$ of the translation-invariant (TI) code in the (2/9,4/9) CDSC family (see Fig.~\ref{fig:TI_code}) on the $L\times L$ square lattice for  $p= 0.2$ and  $\eta = 100$. The subthreshold logical error rates for the XZZX code and the XY code are shown for comparison. Bottom: Thresholds $p_\text{th}$ for the TI code and the hashing bound $p_{\text{h.b.}}$ as functions of bias $\eta$.} 
    \label{fig:TI_code_subthreshold_data}
\end{figure}

We investigate the performance of a particular translation invariant code in this family that is illustrated in Fig.~\ref{fig:TI_code} for size $L=9$.  
The subthreshold logical error rates, using the BSV decoder, at physical error rate $p=0.2$ and bias $\eta=100$ are shown in Fig.~\ref{fig:TI_code_subthreshold_data} (top). The logical error rates for the XZZX code, XY code, and the random CDSC family $(0.25,0.5)$ (average) are shown for comparison. It is interesting to note that this translation-invariant code not only beats the performance of XZZX and XY codes but also beats the average performance of the optimal random CDSC family $(0.25,0.5)$. Thus, this translation-invariant code does not yield the typical performance of the (2/9,4/9) random CDSC family which typically underperforms the (0.25,0.5) random CDSC family. In general, it is possible to design codes in a family that are atypical in code performance.

We find the BSV decoder thresholds of this code to be close to those of the $(0.25,0.5)$ CDSC family and track the Hashing bounds as shown in Fig.~\ref{fig:TI_code_subthreshold_data} (bottom). The detailed plots of logical error rate $p_\text{logical}$ versus physical error rate $p$ and finite-size scaling at different biases $\eta$ are shown in Fig.~\ref{fig:thr_fss_TI_code} in the Appendix.

\section{Discussion}In this work, we introduced the concept of (random) Clifford deformations of QEC codes and used it to construct \mbox{CDSCs}.
We observed that for Pauli noise with finite bias $\eta$, random \mbox{CDSCs} over a broad range of parameters outperform carefully constructed translationally invariant codes, such as the XY and XZZX surface codes.
Thus, the choice of Clifford deformation is an important optimization parameter in QEC that goes beyond the choice of the lattice and its boundaries.
We expect that exploring spatially non-uniform Clifford deformations, quasiperiodic, or translation invariant ones with bigger unit cells will lead to high-performing QEC codes under biased noise.  As an example, we provided an explicit derandomized version of one of our high-performance random CDSC family and showed that the high performance is preserved.

To leverage the benefits of \mbox{CDSCs} into practical universal computation, bias-preserving syndrome measurement circuits, and fault-tolerant logical gates need to be found. 
The statistical mechanics approach to QEC code optimization pursued here should yield similar benefits in the fault-tolerant setting.
Using equivalence relations under global Pauli permutations, our results establish that the $(0.5,\piyz)$ random \mbox{CDSC}s have a 50\% threshold for noise infinitely biased towards any of Pauli $X$, $Y$, or $Z$ errors (similar to the XZZX surface code). We expect such random CDSCs to have favorable performance when the bias direction is non-uniform in space while keeping the bias magnitude large i.e., different qubits can be either $X$-biased, $Y$-biased, or $Z$-biased. This is more clear if we work in the Heisenberg picture, in which we can consider the Clifford deformation on the noise instead of the stabilizers. Our phase diagram then illustrates spatially non-uniform configurations of noise that yield high performance from the CSS surface code. Now, consider a random CDSC family, let's say (0.25,0.5), and a noise model with large bias and spatially non-uniform bias directions, instead of the uniform $Z$-biased noise. In the Heisenberg picture, this effectively yields a spatially non-uniform noise model acting on the CSS code, that is different from the one given by the parameters (0.25,0.5); in other words, this leads to new effective parameters $(\Pi_{XZ}^{\text{eff}},\Pi_{YZ}^{\text{eff}})$. For a wide class of spatially non-uniform configurations of bias directions, these new effective parameters $(\Pi_{XZ}^{\text{eff}},\Pi_{YZ}^{\text{eff}})$ are expected to lie in our high-threshold phase and yield high performance. Overall, CDSCs may provide an attractive approach to building scalable quantum computers.


\section*{Acknowledgements}We thank David Huse, Grace Sommers, and Giacomo Torlai for valuable discussions, Ali Lavasani and Ben Criger for useful feedback on the manuscript and numerics, Eric Huang for sharing his plotting routines, and Victor Albert for help with errorcorrectionzoo.org entry on CDSCs~\cite{CDSCeczoo}. The tensor network decoder for the random \mbox{CDSCs} was implemented via an appropriate adaptation of~\textit{qecsim} \cite{qecsim} (a Python3 package for simulating quantum error correction using stabilizer codes) and simulations were run on the NIST Raritan cluster. The simulation code and data are available on request. We acknowledge support from the National Science Foundation (QLCI grant OMA-2120757), ARO (W911NF-18-1-0212), the Packard Foundation (2020-71479), the Simons Foundation through the collaboration on Ultra-Quantum Matter (651438, AD) and by the Institute for Quantum Information and Matter, an NSF Physics Frontiers Center (PHY-1733907).
\bibliographystyle{apsrev4-2}
\bibliography{bib}

\begin{thebibliography}{49}
\bibitem [{ran()}]{random}%
  \BibitemOpen
  \href@noop {} {}\bibinfo {note} {Note, random families of surface codes have also been recently studied and found to have favorable performance in other contexts \cite{Li19,Chubb21}.}\BibitemShut {Stop}%
\end{thebibliography}

\appendix

\section{Adaptation of the BSV decoder and simulations}
\label{app:BSV}
The Bravyi-Suchara-Vargo (BSV) decoder~\cite{BSV2014} is an efficient approximation of the optimal maximum likelihood (ML) decoder~\cite{Dennis2002}. 
The ML decoder achieves the same error threshold as that obtained from the critical point of the statistical-mechanical mapping of error correction. The BSV decoder achieves a value close to it. Given a particular syndrome configuration, the goal is to calculate the probabilities of logical error classes and provide a correction operator based on the logical class with the maximum probability. 
Exact evaluation of the logical error class probabilities is, in general, inefficient. 
The BSV decoder uses an algorithm that efficiently approximates the logical error class probabilities through tensor-network contractions. 
The size of the tensors being contracted is reduced through Schmidt decomposition and retaining only the $\chi$ largest Schmidt values, \textit{i.e.}, $\chi$ is the bond dimension of the tensors. 
For a modest value of $\chi$, the method converges for a range of noise models and boundary conditions of the surface code. 

\begin{figure}[t]
\centering
\includegraphics[scale=0.14]{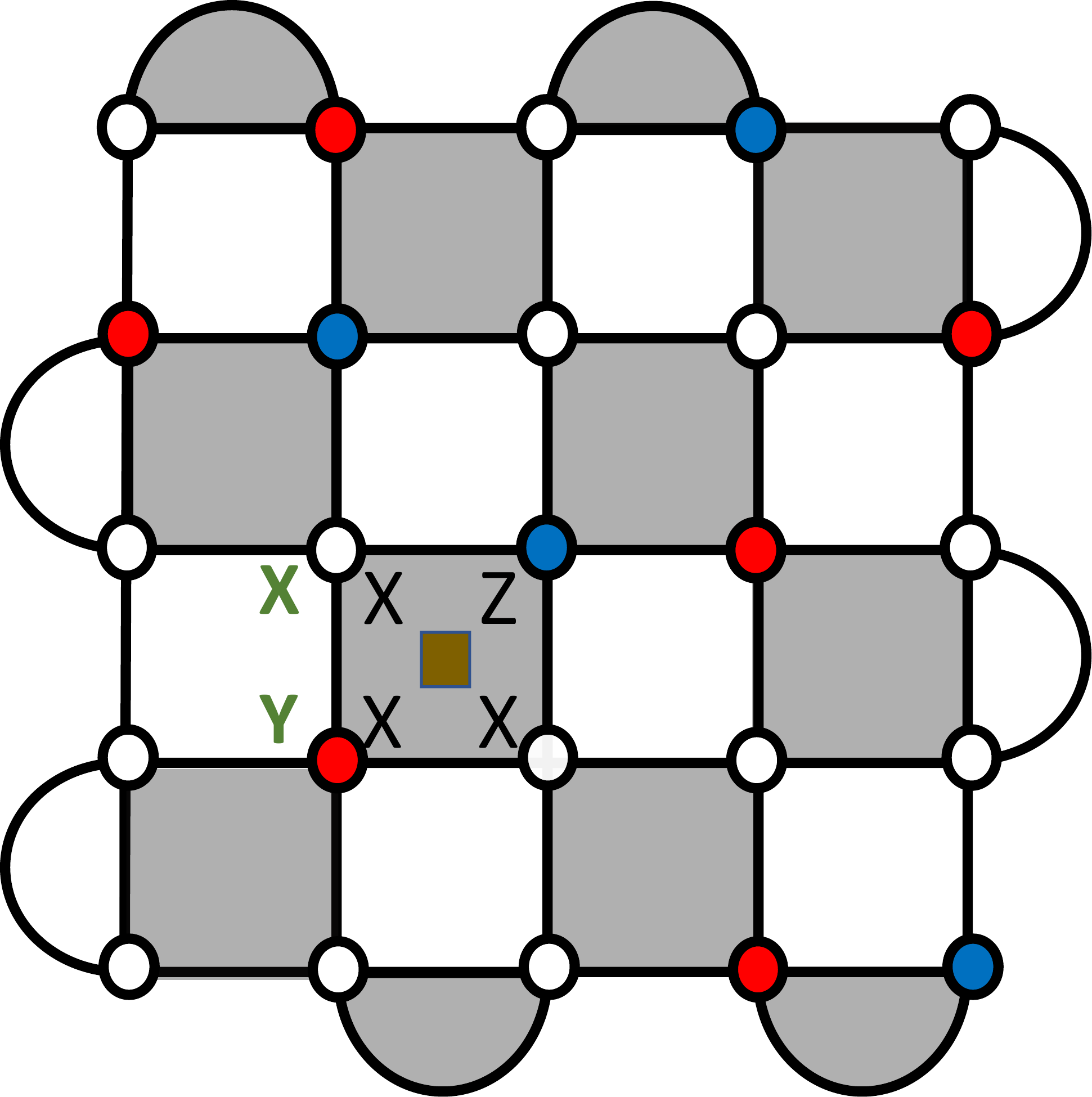}
\caption{Example to describe the modification in the evaluation of logical error class probability due to Clifford deformations. Pauli error is shown in green letters. The goal is to evaluate the probability of the product of the error times the marked stabilizer; this product appears as a term in the logical error class probability evaluated via tensor network contraction. Without the Clifford deformations, error times the marked stabilizer gives two $X$ operators and one $Z$ operator resulting in a probability $p_X^2 p_Z (1-p)^{L^2-3}$ where $p$ is the total error rate and $L^2$ is the number of qubits. Because of Hadamard $\Hadamard$ (blue) and $H_{YZ}=\Hadamard\sqrt{Z}\Hadamard$ (red) Clifford deformations acting on two qubits of the marked stabilizer, we instead get three $X$ operators resulting in probability $p_X p_Z^2(1-p)^{L^2-3}$.
Equivalently, we can modify the local noise model on the qubits, according to the Clifford deformations while keeping the stabilizers same as the CSS surface code to get the same logical error class probability as with deformed stabilizers and a spatially uniform $Z$ bias model.}
\label{tensornetworkmodification}
\end{figure}

Our numerical implementation uses a few minor modifications to the BSV decoder for the standard form of the surface code with a uniform noise model. To make these modifications, we keep the Heisenberg picture in mind, \textit{i.e.}, we keep the standard form of the stabilizers but have a spatially varying noise model depending on the Clifford deformations that are applied to the qubits. In the Heisenberg picture, the tensor redefinitions can be understood as having a different local noise model on each qubit leading to the redefinition of the associated tensor. For the random code with a given ratio of Clifford deformations $\pixz$ and $\piyz$, we modify the noise model on each tensor associated with a qubit according to the Pauli permutation acting on it. 
For example, a local noise model given by $(p_X,p_Y,p_Z)$ maps to $(p_Z,p_Y,p_X)$ under the Hadamard deformation $\Hadamard$ acting on the qubit. We illustrate this for a particular code realization in Fig.~\ref{tensornetworkmodification}.

\section{Cluster method for Maximum Likelihood thresholds}
\label{app:cluster_summation}
In this appendix, we describe the approximate analytical method for calculating the critical points of random-bond Ising models (RBIMs) that are self-dual in the absence of disorder~\cite{Nishimori_Ohzeki_Nihat_2008,Ohzeki_2009}. We use this approach to estimate the optimal thresholds for the XY code. The partition function of the RBIM
is given by 
\begin{align}
    Z=\sum_{\{s_i\}}\prod_{\{i\}}\text{LBF}(\{s_i\},\{\tau_\text{P} J_\text{P}\})
\end{align}
where $\text{LBF}(\{s_i\},\{\tau_\text{P} J_\text{P}\})$ is the local Boltzmann factor (LBF) involving spins $\{s_i\}$ at positions $\{i\}$ in a local neighborhood, interacting with coupling coefficients $J_\text{P}$ that have sign disorder $\tau_\text{P}$. The statistical mechanical model in Eq.~(3) of main text is the so-called 8-vertex model. The Hamiltonian (neglecting boundary terms) can be written more explicitly as

\begin{eqnarray}
\mathcal H = &-& \sum_{\figbox{.12}{eq_summand}}
\tau_X J_X\figbox{.2}{eq_szz}
+\tau_Y J_Y\figbox{.2}{eq_sxxzz}
+\tau_Z J_Z\figbox{.2}{eq_sxx}\nonumber\\
&-&\sum_{\figbox{.12}{eq_summand_rot}}
\tau_X J_X\figbox{.2}{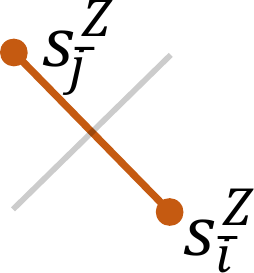}
+\tau_Y J_Y\figbox{.2}{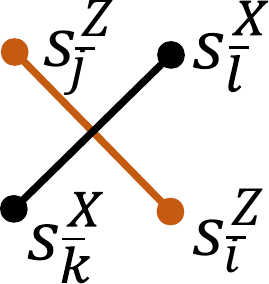}
+\tau_Z J_Z\figbox{.2}{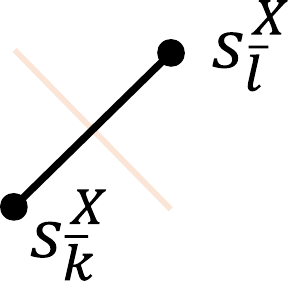},\quad\quad
\label{eqn:SM_Hmain_explicit}
\end{eqnarray}
where $s_*^X$ and $s_*^Z$ are Ising spins associated with every $X$ and $Z$ stabilizer generator of the surface code, and the summation is over all qubit locations, \textit{i.e.}, $\figbox{.12}{eq_summand}$ and $\figbox{.12}{eq_summand_rot}$ crossings; see Fig.~3(b) in the main text.

For the 8-vertex model, the $\text{LBF}(\{s_i\},\{\tau_\text{P} J_\text{P}\})$ at inverse temperature $\beta$ for the spins $\{s_i\}$ on a $\figbox{.12}{eq_summand}$ is given by
\begin{align}
    &\text{LBF}(\{s_i\},\{\tau_\text{P} J_\text{P}\})=\nonumber\\
&\exp\Bigg(\tau_X J_X\figbox{.2}{eq_szz}
+\tau_Y J_Y\figbox{.2}{eq_sxxzz}
+\tau_Z J_Z\figbox{.2}{eq_sxx}\Bigg).
\label{eqnLBF}
\end{align} The LBF associated with neighboring qubit of the underlying surface code is similar but with the colors switched, \textit{i.e.}, $\figbox{.12}{eq_summand_rot}$. To average over the sign disorders $\tau_\text{P}$ in coupling coefficients $J_\text{P}$, we consider a replicated system with $n$ replicas, whose partition function we can write as follows,
\begin{align}
    Z_n=\langle\big[\sum_{\{s_i\}}\prod_{\{i\}}\text{LBF}(\{s_i\},\{\tau_\text{P} J_\text{P}\})\big]^n\rangle. 
\end{align}
Here, $\langle...\rangle$ denotes an average over configurations with different disorder realizations. The LBF of the replicated system can be written as follows,
\begin{align}
    \text{LBF}^{(n)}(\{s_i\}\{J_\text{P}\})=\langle\big[\prod_{\{i\}}\text{LBF}(\{s_i\},\{\tau_\text{P} J_\text{P}\})\big]^n\rangle. 
\end{align}
We now consider an $n$-binary Fourier transform $Z^\star$ of the partition function of the replicated system as follows,
\begin{align}
    Z^\star_n=\sum_{\{s_i\}}\prod_{\{i\}}\text{DBF}^{(n)}(\{s_i\},\{J_\text{P}\})
\end{align}
where $s_i$ indicates the spins in a dual lattice and $\text{DBF}^{(n)}(\{s_i\},n,\{J_\text{P}\})$ refers to the dual Boltzmann factor which is given by the $n$-binary Fourier transformation of the LBF. When the coupling constants $J_\text{P}$ are such that the model without quenched disorder is self-dual, then we can equate the partition function and its $n$-binary Fourier transformation to get a good approximation to the critical point of the disordered statistical-mechanical model. This critical point is an approximation of the ML threshold of the error correcting code. The product in the definition of $Z_n$ and $Z^\star_n$ is hard to evaluate over the full lattice and can be approximated by taking it over a cluster of spins. The required cluster size increases for a better approximation to the threshold. We show the clusters we used to get the thresholds in Fig.~3(b) of the main text.  

\begin{figure}[t!]
\centering
\vspace{6mm}
\includegraphics[width=0.99\textwidth]{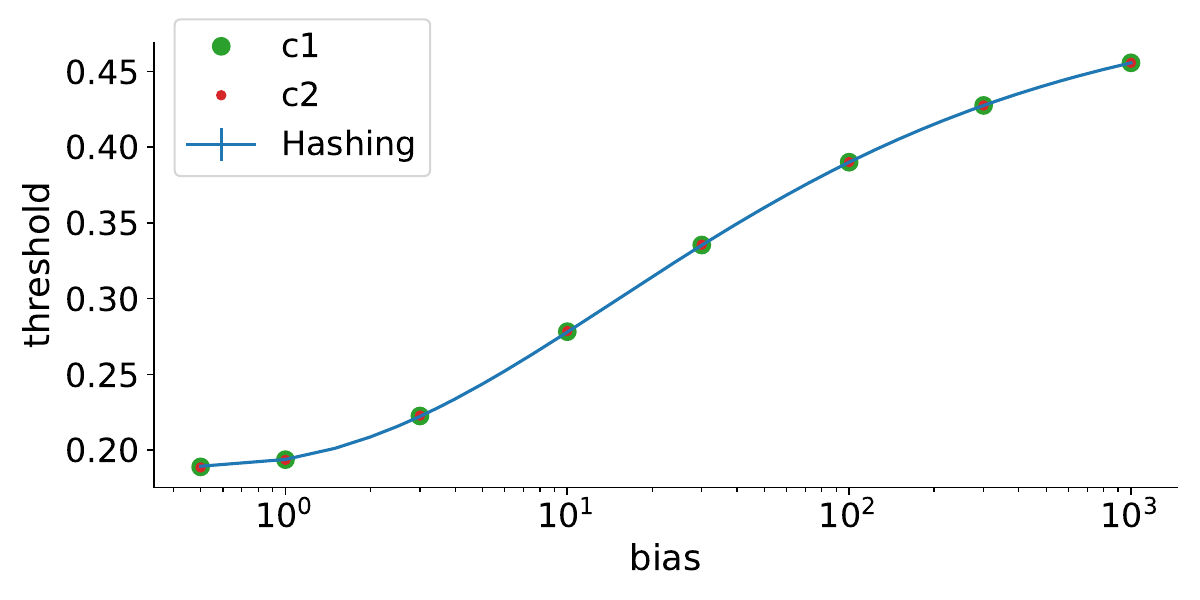}
\includegraphics[width=0.99\textwidth]{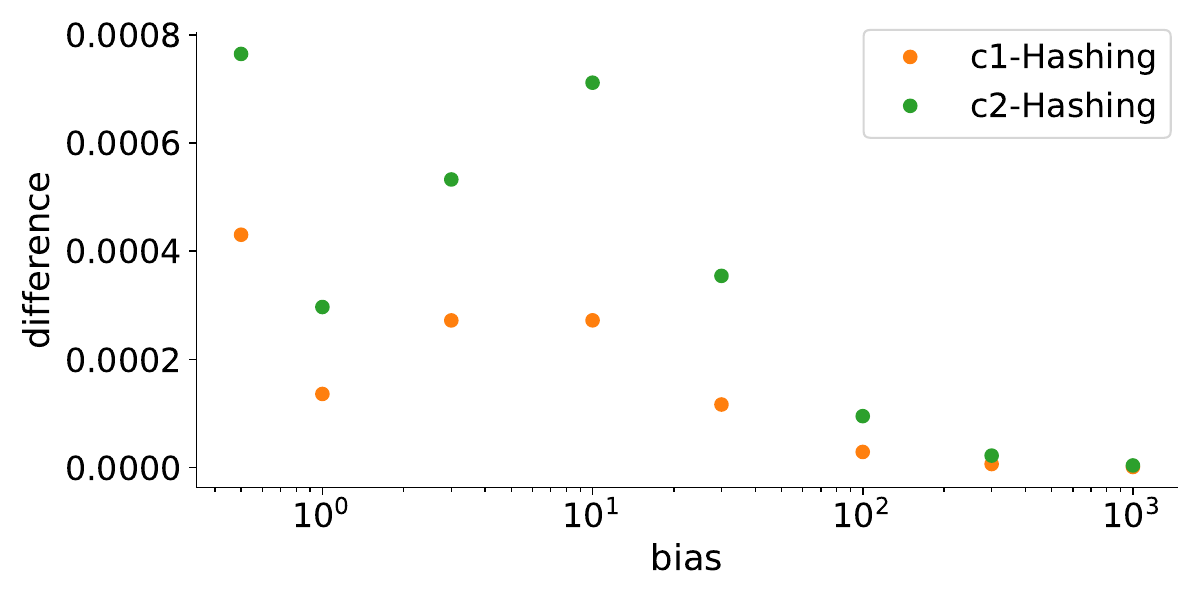}

\caption{Top: Cluster-based approximations to the finite-bias threshold track the Hashing bound.  $c=0$ approximation is exactly the same as the hashing bound, \textit{i.e.}, $p_{\text{th},c=0}=p_\text{h.b.}$ and hence, not shown separately.
$c=1,2$ approximations are shown using green and red markers with legends c1, c2 respectively. 
Bottom: The difference between the threshold values from $c=1,2$ cluster approximations and the hashing bound, i.e., $p_\text{th}-p_\text{h.b.}$ is shown.}
\label{fig:thresholds_Ohzeki}
\end{figure} 

We consider the limit $n\rightarrow 1$ to get a good approximation to the critical point of the disordered statistical mechanical model. In this limit, $Z_n=Z^\star_n$ reduces to the following equation,
\begin{align}
    \langle\hspace{1mm}\log\sum_{\{s_f\}}\prod_{\text{cluster}}{\text{LBF}}\hspace{1mm}\rangle=\langle\hspace{1mm}\log\sum_{\{s_f\}}\prod_{\text{cluster}}{\text{DBF}}\hspace{1mm}\rangle,
    \label{eqn:thr_selfduality}
\end{align}
where $\langle ...\rangle$ denotes the disorder average. The 8-vertex model is self-dual when the coupling coefficients $J_X=J_Z$. The LBF expressed in Eq.~\eqref{eqnLBF} can be written as
\begin{align*}
    &\text{LBF}\\
    =&\exp\Big[\tau_X J_X s^Z_1 s^Z_2+ \tau_Z J_Z s^X_{1}s^X_{2}+
    \tau_X\tau_Z J_Y s^X_1 s^X_2 s^Z_{1}s^Z_{2}\Big],
\end{align*}
where $s^X$ denote spins corresponding to the $X$ stabilizers (at the ends of the black bond in Eq.~\eqref{eqnLBF}) and $s^Z$ denote spins corresponding to the $Z$ stabilizers (at the ends of the orange bond in Eq.~\eqref{eqnLBF}). Taking the Fourier transform, we get the dual Boltzmann factor DBF as follows,
\begin{align*}
    & \text{DBF}\\
    =&\frac{1}{2}\exp\Big[\tau_X J_X+ \tau_Z J_Z+\tau_X\tau_Z J_Y\Big] \\
    +&\exp\Big[\tau_X J_X- \tau_Z J_Z-\tau_X\tau_Z J_Y\Big] s^X_{1}s^X_{2}\\
    +&\exp\Big[-\tau_X J_X+ \tau_Z J_Z-\tau_X\tau_Z J_Y\Big]  s^Z_1 s^Z_2\\
    +&\exp\Big[-\tau_X J_X-\tau_Z J_Z+\tau_X\tau_Z J_Y\Big]  s^X_1 s^X_2 s^Z_{1}s^Z_{2}.
\end{align*}
$c=0$ result is same as the hashing bound $p_\text{h.b.}$ curve in  Fig.~\ref{fig:thresholds_Ohzeki}. We obtained the thresholds $p_{\text{th},c=1}$ and $p_{\text{th},c=2}$ for the XY code using the larger clusters $c=1$ and $c=2$ at different values of $Z$ biases $\eta$. The difference of the thresholds $p_\text{th}$ from the hashing bound $p_\text{h.b.}$ is also shown in Fig.~\ref{fig:thresholds_Ohzeki}. 
\vspace{5mm}
\section{Subthreshold scaling plots}
\label{app:subthr_scaling}

In Fig.~\ref{fig:all_plots_subthreshold_scaling} below, we show the plots depicting subthreshold performance of the random CDSCs, the XY code and the XZZX code at different biases $\eta=10,100,1000$ and physical error rate $p=0.2$. We calculated the logical error rate from 200 000 Monte Carlo runs for code distances $L=9,13,17,21$. Each Monte Carlo run for simulating error correction on the random CDSC is done on a different realization consistent with the probabilities $(\pixz,\piyz)$. Error bars were calculated through jackknife resampling. 
We also include the XZZX on a torus with coprime dimensions $L\times (L+1)$ using a minimum-weight perfect matching (MWPM) decoder for comparison since it has been shown to have extraordinary subthreshold performance as well. However, since the decoder, the number of qubits ($L(L+1)$ as opposed to $L^2$ for the square lattice codes) and the boundary conditions are all different, the comparison's goal is only to justify that the random CDSCs can beat the best known subthreshold performance.

\begin{figure*}[t]
  \begin{center}
  \includegraphics[scale=1.5]{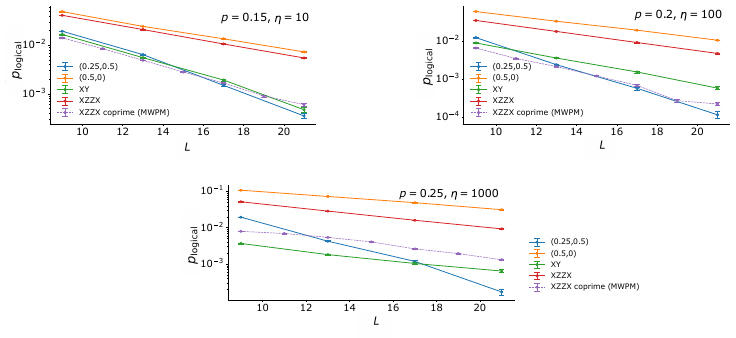}
     \caption{ Sub-threshold logical error rates of various \mbox{CDSCs} on the $L\times L$ square lattice for $p= 0.2$ and  $\eta = 10,100,1000$. The performance of the XZZX code on a $L\times (L+1)$ torus~\cite{XZZX2021} using the MWPM decoder is included for comparison.}
    \label{fig:all_plots_subthreshold_scaling}
  \end{center}
\end{figure*}

\section{Performance of random CDSCs and translation-invariant code realization}
\label{app:finitebias_thr_plots}
In Figs.~\ref{fig:thrandfss_randomcodes} and \ref{fig:thrandfss_randomcodes2} below, we plot the logical error rates of the random CDSCs $(0.25,0.5)$ and $(0.5,0.0)$ at different biases $\eta=10,30,100,1000$. 
The logical error rate for a particular random CDSC is calculated using $120\, 000$ such runs using the BSV decoder for a set of physical error rates near the threshold $p_c$ for code distances $L \in \{$9, 13, 17, 21$\}$. 
For these simulations, we used a large value of the bond dimension, $\chi = 56$. 
We observe that the decoder converges close to the threshold for this bond dimension at lower biases such as $0.5$ and $10$. 
However, close to the threshold at biases of $100$ and $1000$, we do not observe complete convergence. Hence, we confirm our threshold values by checking the exponential decay of the logical error rate at rates below the threshold. 

In Fig.~\ref{fig:thr_fss_TI_code}, we show the logical error rates of the translation-invariant code from the random CDSC family $(2/9,4/9)$ at moderate biases $\eta=10,30,100,1000$. We observe that the thresholds track those of the random families discussed above; the threshold numbers are plotted in Fig.~\ref{fig:TI_code}(bottom) in the main text. 

We also show the finite size scaling plots where we used the critical exponent method~\cite{Wang2003}, \textit{i.e.}, the values of thresholds were obtained from the logical error rate data by doing finite-size scaling analysis, \textit{i.e.}, fitting the failure rate curves to the logical error rate  $p_\text{logical}=A+Bx+Cx^2$, where
$A$, $B$, $C$, $p_\text{th}$ and $\nu$ are fitting coefficients, $x=(p-p_\text{th})L^{1/\nu}$ and $p$ is the error rate.
Note that $p_\text{th}$ and $\nu$ correspond to the threshold and critical exponent, respectively.

\vspace{4mm}
\begin{figure*}[h]
  \begin{center}
  \includegraphics[scale=1.5]{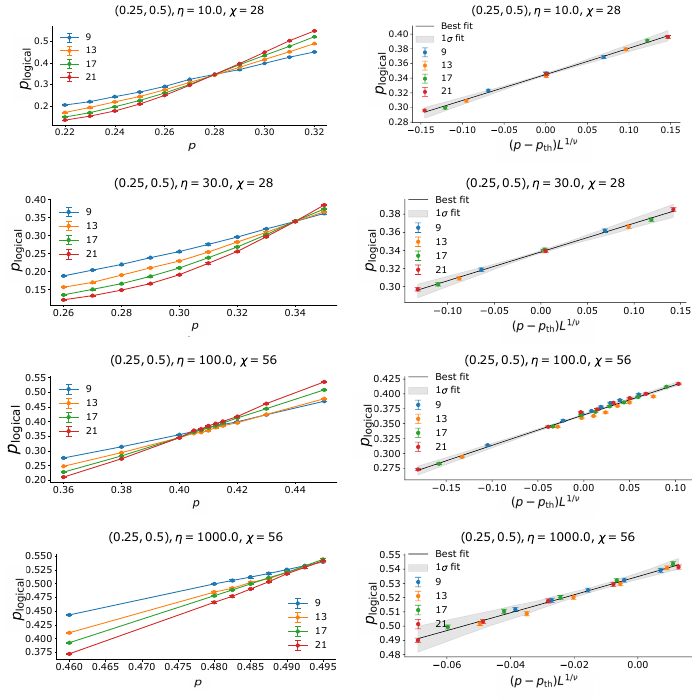}

  \caption{Logical error rate $p_\text{logical}$ versus physical error rate $p$ and finite size scaling for the (0.25,0.5) random CDSC at different biases $\eta$ and using bond dimension $\chi$ for the BSV decoder as indicated. Each data point for logical error rate $p_\text{logical}$ is averaged over 120 000 Monte Carlo runs.}
    \label{fig:thrandfss_randomcodes}
  \end{center}
\end{figure*}

\begin{figure*}
  \begin{center}
        \includegraphics[scale=1.5]{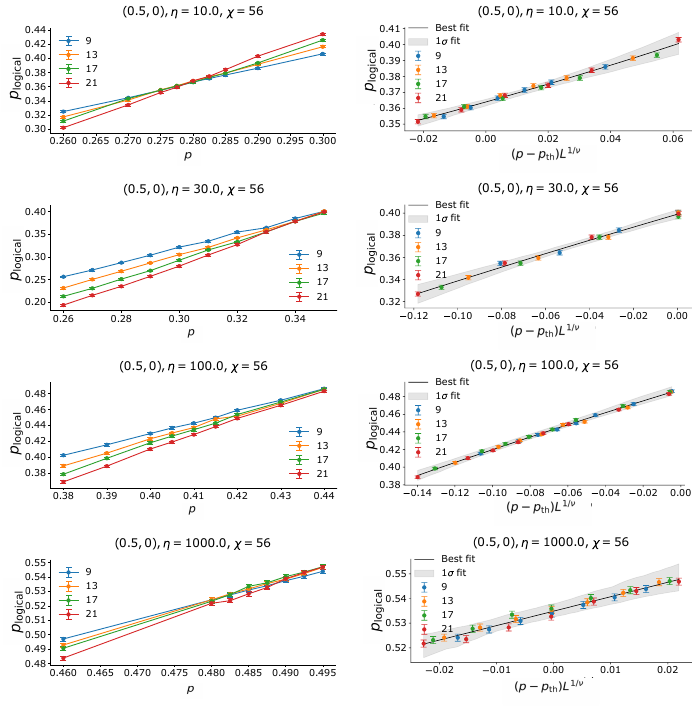}

     \caption{Logical error rate $p_\text{logical}$ versus physical error rate $p$ and finite size scaling for the (0.5,0) random CDSC at different biases $\eta$ and using bond dimension $\chi$ for the BSV decoder as indicated. Each data point for logical error rate $p_\text{logical}$ is averaged over 120 000 Monte Carlo runs.}
    \label{fig:thrandfss_randomcodes2}
  \end{center}
\end{figure*}

\section{Random CDSCs at large bias \texorpdfstring{$\eta=10^8$}{eta=1e8}}
\label{app:random_thr_plots}
\begin{figure*}[h]
  \begin{center}
 \raisebox{3mm}{\normalsize  \textbf{(a).} \par}\\  \includegraphics[scale=0.15]{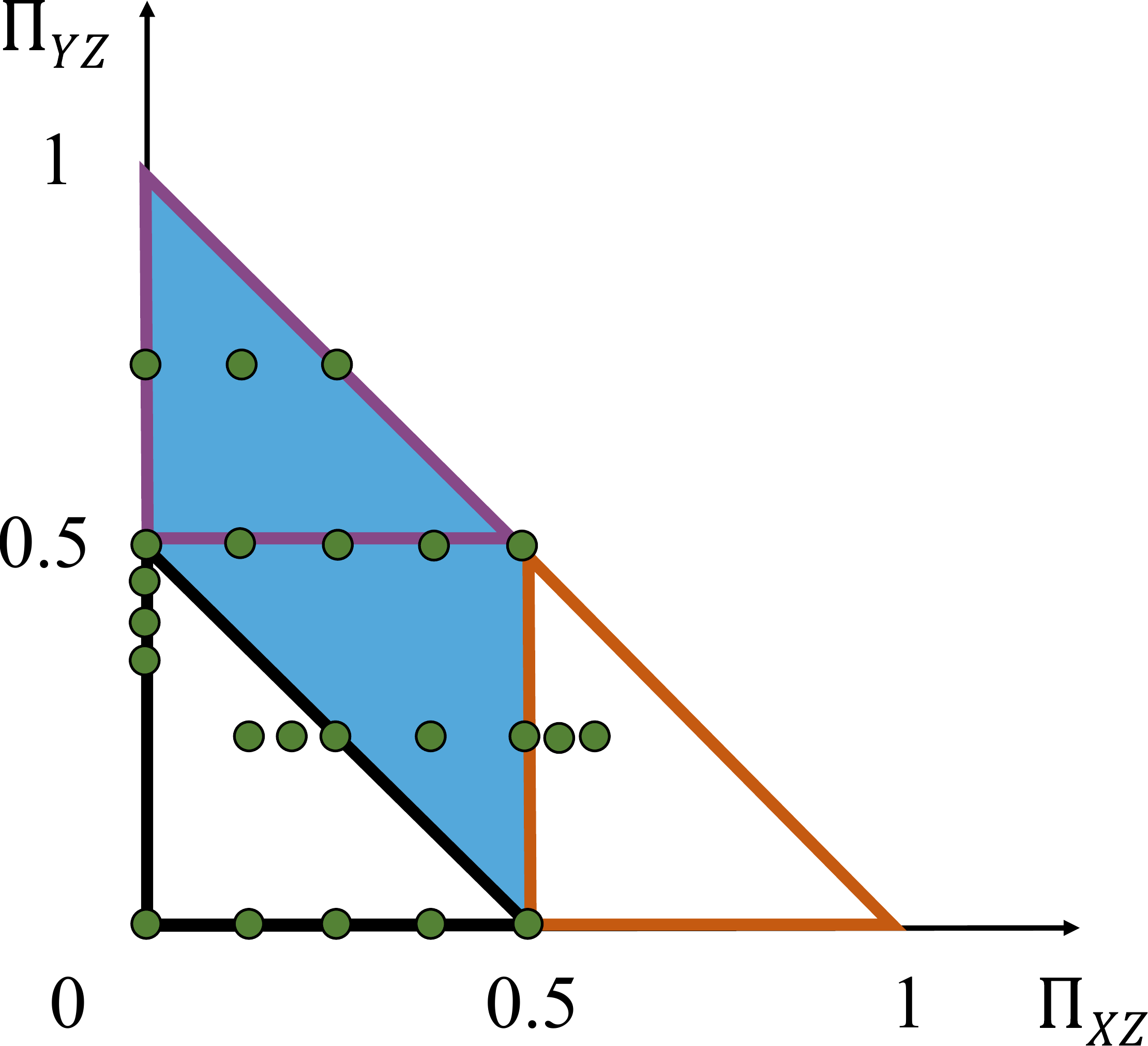}\\
 \vspace{5mm}
 \raisebox{3mm}{\normalsize  \textbf{(b).} \par} \\
 \vspace{2mm}
 \includegraphics[scale=1.5]{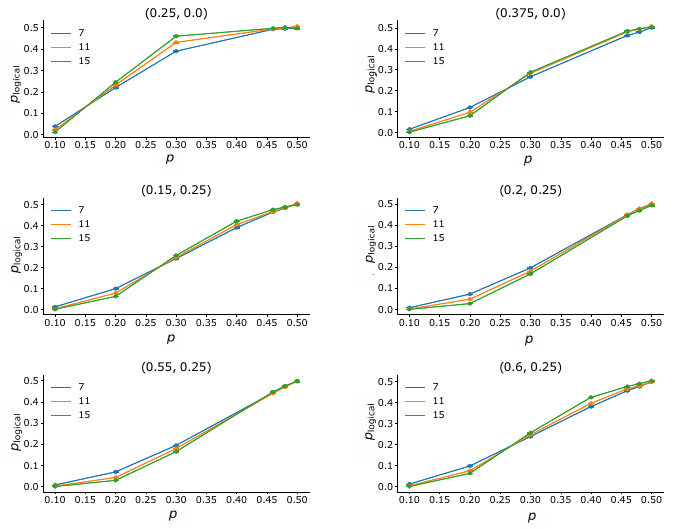}
  \caption{(a) 50\% threshold phase diagram. The random CDSCs whose code performance was studied numerically are marked (green circles). (b) Performance of Random CDSCs outside the blue region (50\% thresholds) of the phase diagram. Logical error rate $p_\text{logical}$ versus physical error rate $p$ for the $(\pixz,\piyz)$ random CDSCs at bias $\eta=10^8$. Each data point is averaged over 60 000 Monte Carlo runs of the BSV decoder with a bond dimension, $\chi= 56$.}

   \label{fig:random_phase_thresholds1}
  \end{center}
\end{figure*}

We tested our conjectured phase diagram of 50\% thresholds for random CDSCs, shown in Fig.~3(a) of main text, via tensor network numerics on a subset of random CDSCs, marked using green circles in Fig.~\ref{fig:random_phase_thresholds1}(a) below. 

\begin{figure*}
\begin{center}

\includegraphics[scale=1.5]{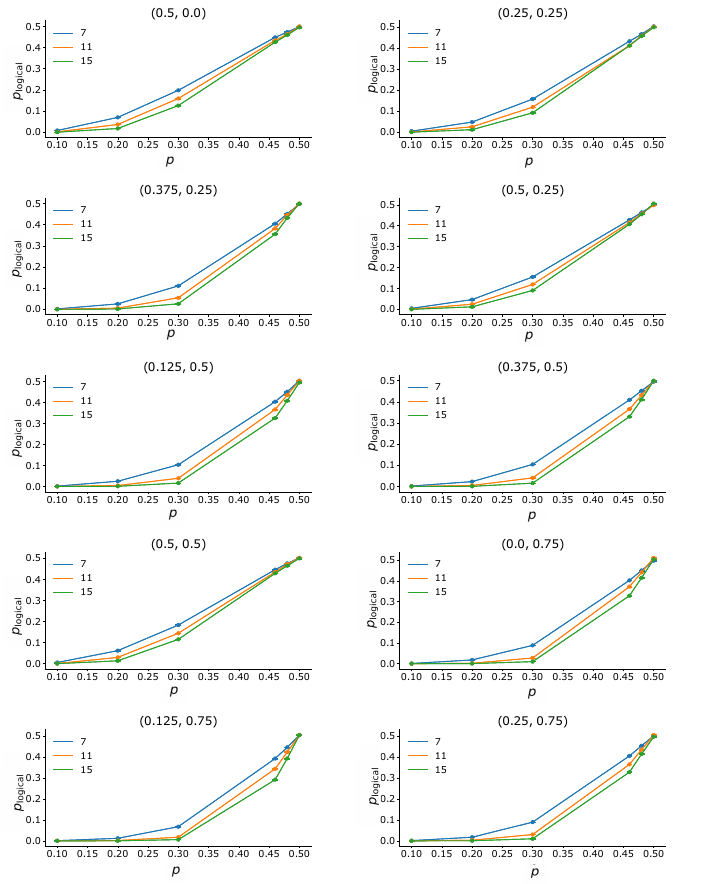}

\caption{Performance of Random CDSCs inside the blue region (50\% thresholds) of the phase diagram. Logical error rate $p_\text{logical}$ versus physical error rate $p$ for the $(\pixz,\piyz)$ random CDSCs at bias $\eta=10^8$. Each data point is averaged over 60 000 Monte Carlo runs of the BSV decoder with a bond dimension, $\chi= 56$.}
\label{fig:random_phase_thresholds2}
\end{center}
\end{figure*}

\begin{figure*}
  \begin{center}
  \vspace{-5mm}
        \includegraphics[height=0.24\textwidth]{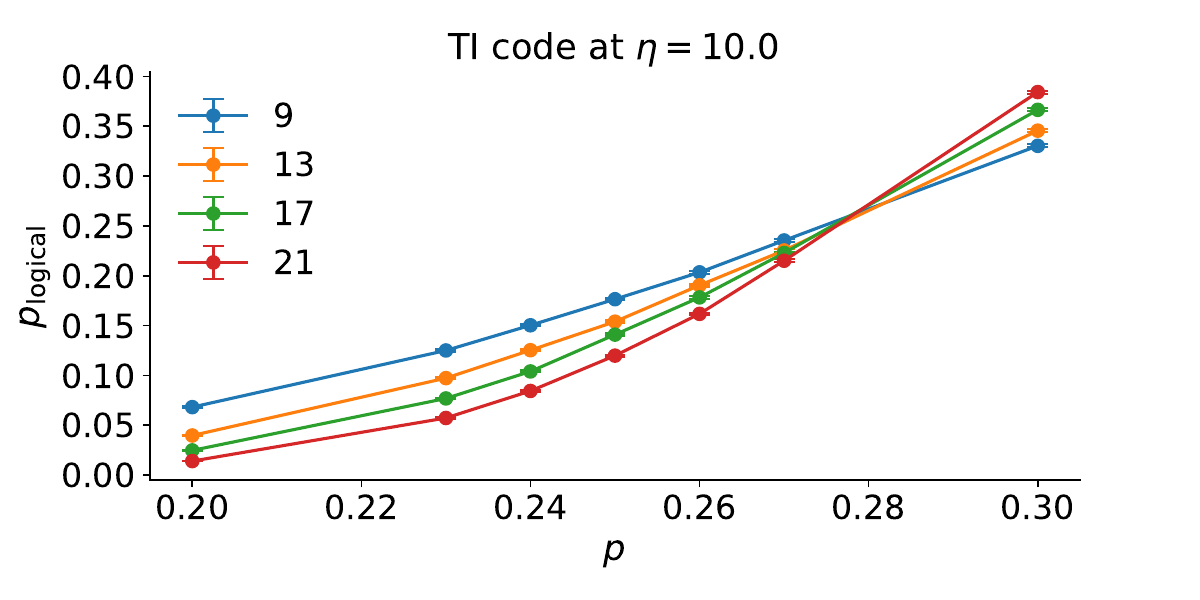}
  \includegraphics[height=0.24\textwidth]{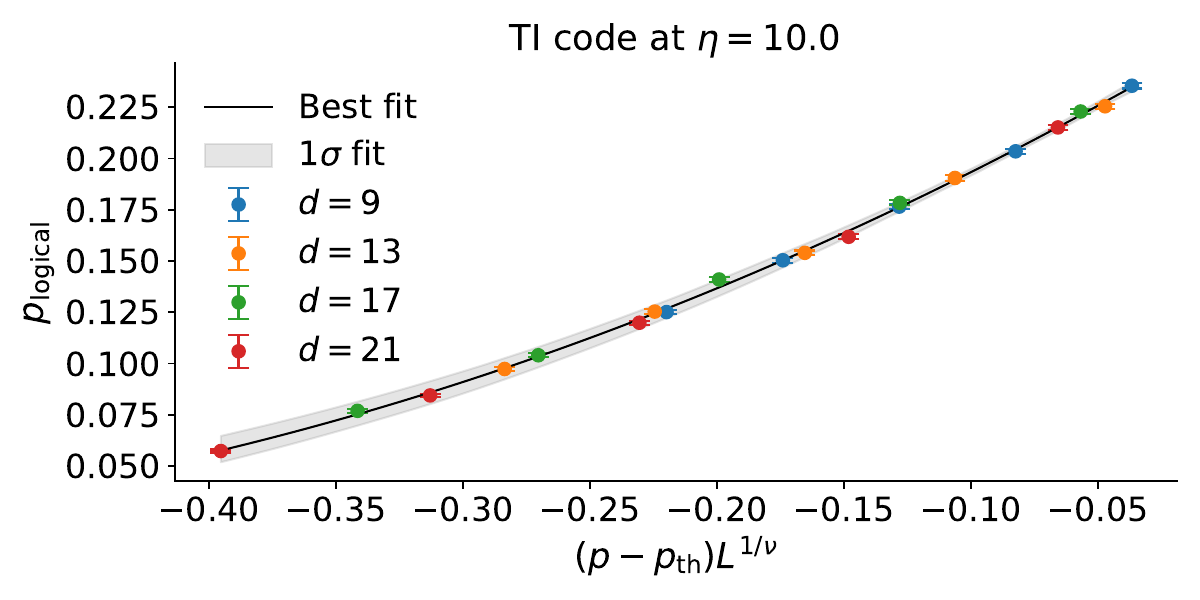}\\
        \includegraphics[height=0.24\textwidth]{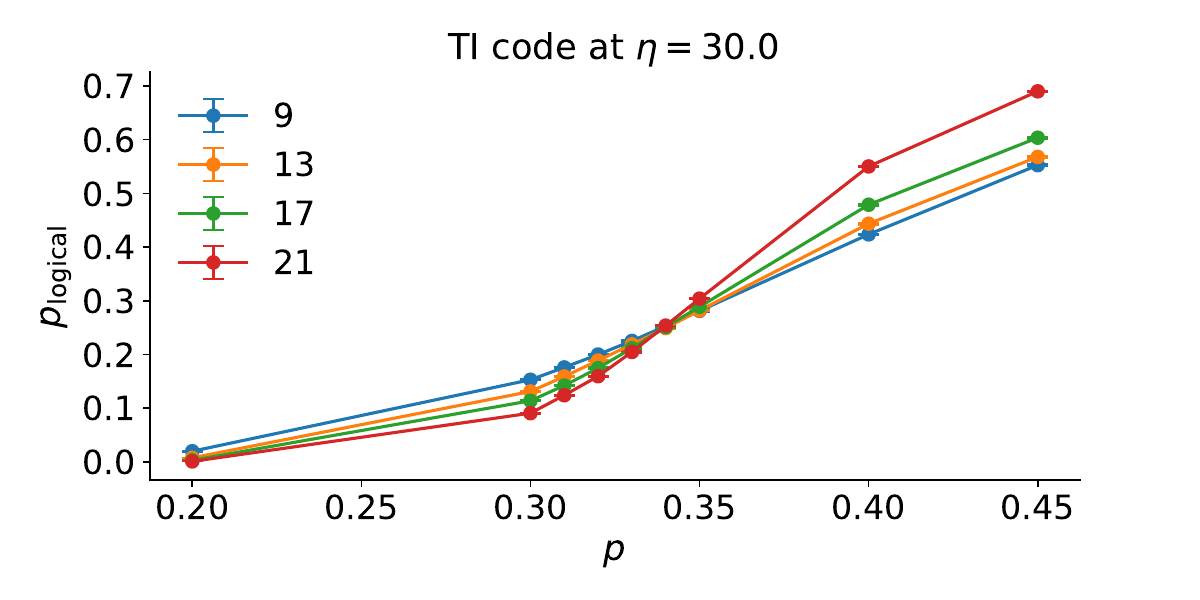}
  \includegraphics[height=0.24\textwidth]{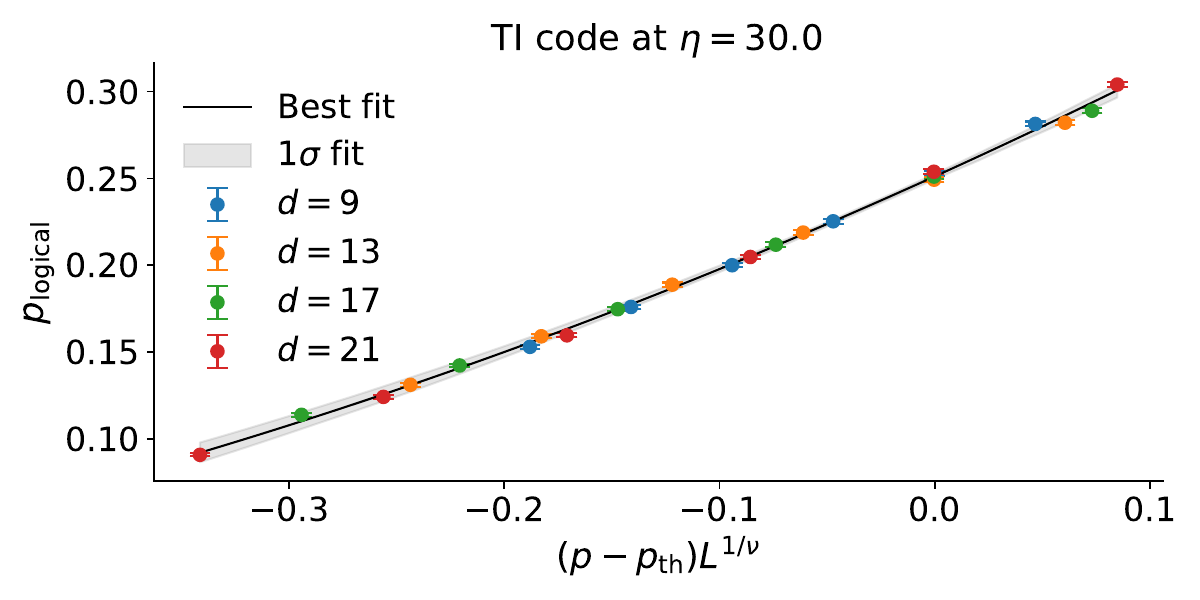}\\

  \includegraphics[height=0.24\textwidth]{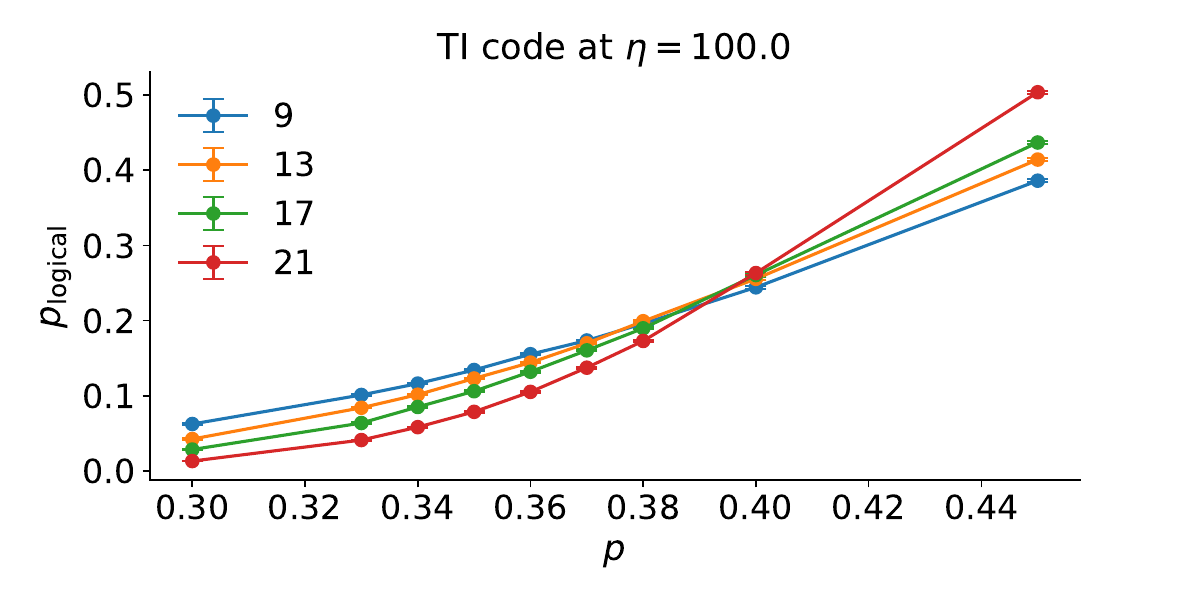}
  \includegraphics[height=0.24\textwidth]{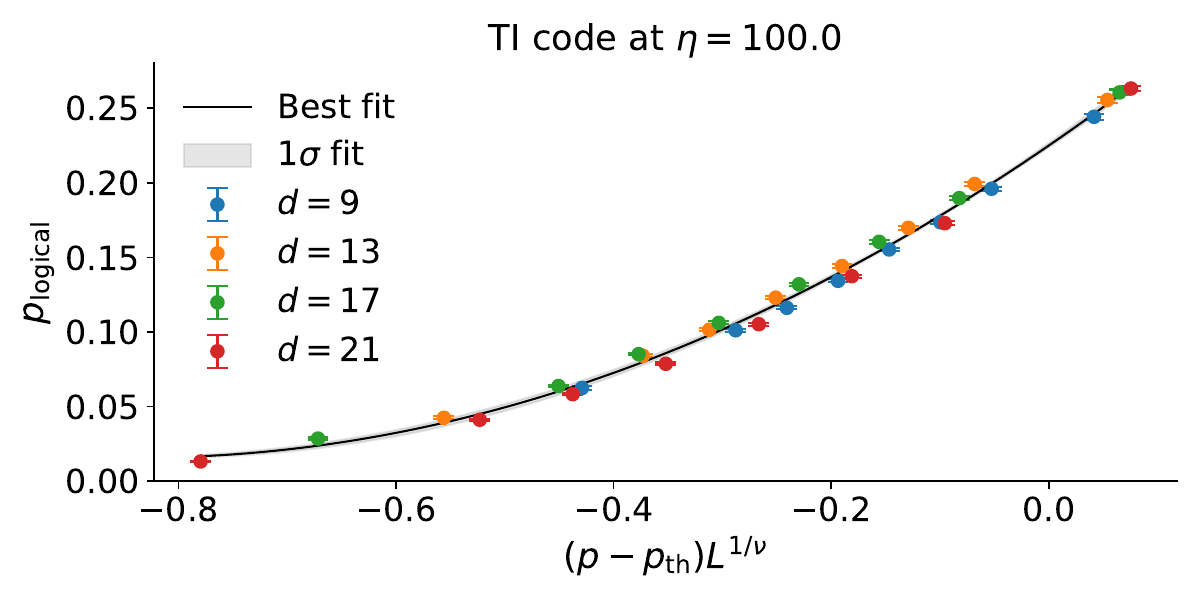}\\

  \includegraphics[height=0.24\textwidth]{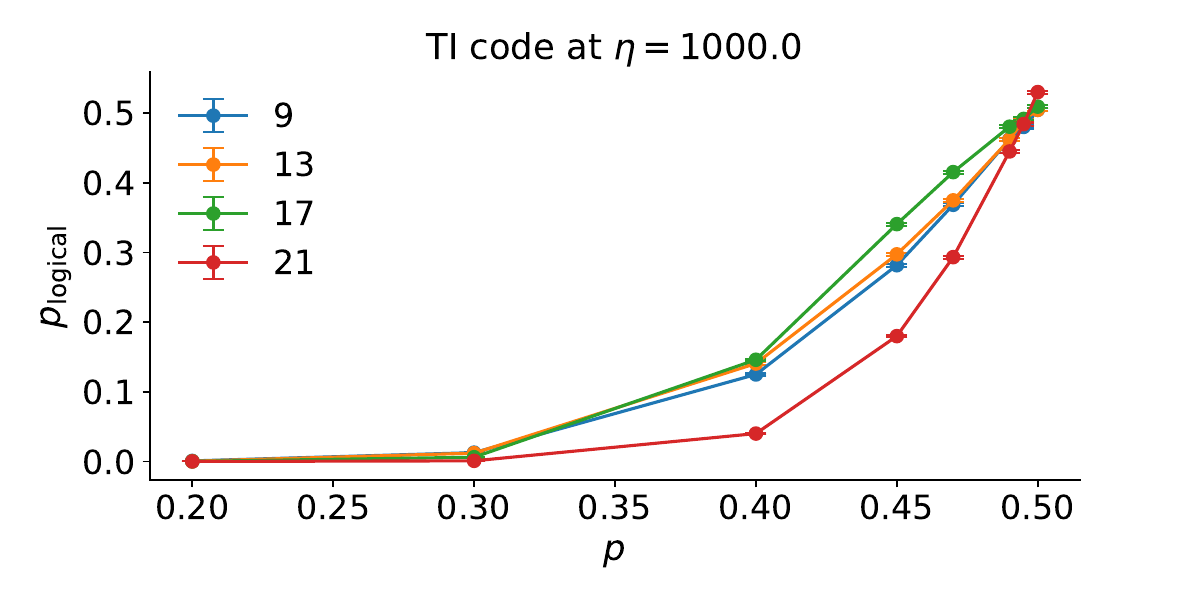}
  \includegraphics[height=0.24\textwidth]{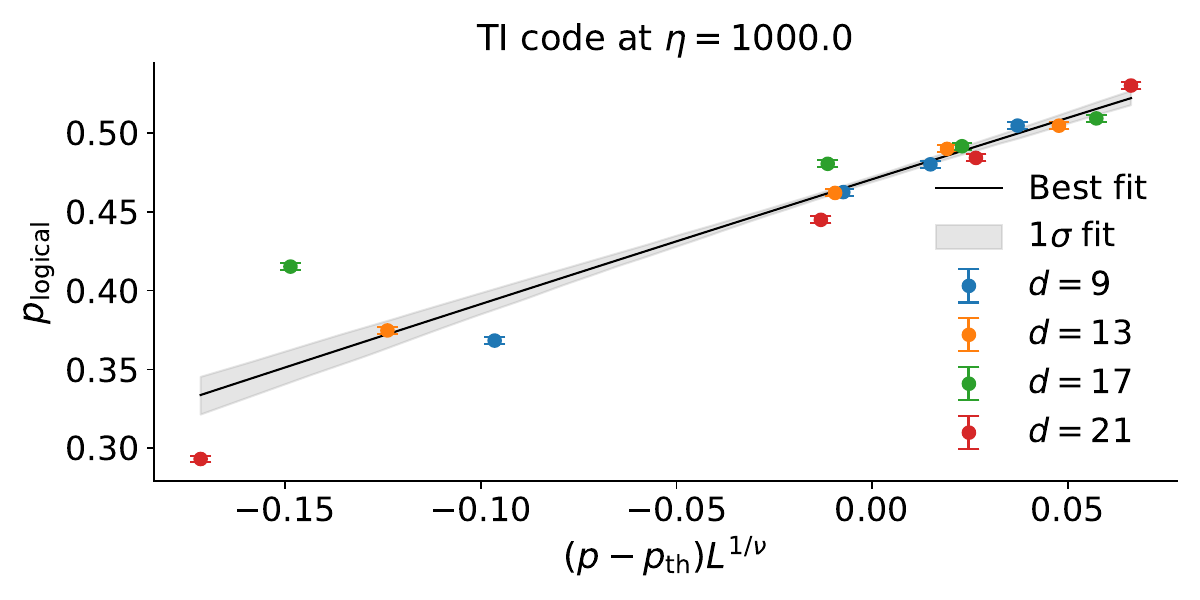}\\

  \vspace{-4mm}
     \caption{Logical error rate $p_\text{logical}$ versus physical error rate $p$ and finite-size scaling for the translation-invariant (TI) code in the CDSC family (2/9,4/9) at different biases $\eta$ and using bond dimension $\chi=56$ for the BSV decoder as indicated. Each data point for logical error rate $p_\text{logical}$ is averaged over 100 000 Monte Carlo runs.}
    \label{fig:thr_fss_TI_code}
  \end{center}
\end{figure*}

In Figs.~\ref{fig:random_phase_thresholds1}(b) and \ref{fig:random_phase_thresholds2}, we show the logical error rates vs physical error rates, obtained using the adaptation of the BSV decoder, for some representative random CDSCs characterized by $(\pixz,\piyz)$ at a bias of $\eta=10^8$. The numerical results are consistent with the infinite bias phase diagram for random CDSCs.

\end{document}